\documentclass[lettersize,journal]{IEEEtran}
\usepackage{cite}
\usepackage{bm}
\usepackage{amsmath}
\usepackage{mathrsfs}

\usepackage{algorithmic}

\usepackage{array}
\usepackage{colortbl}

\ifCLASSOPTIONcompsoc
\usepackage[caption=false,font=normalsize,labelfont=sf,textfont=sf]{\part{title}subfig}
\else
\usepackage{subfigure}
\fi
\usepackage{url}
\usepackage{graphicx,amsmath,amssymb,amsfonts}
\usepackage{algorithmic,algorithm}
\usepackage{stfloats}
\usepackage{pifont}
\usepackage{amsmath}
\usepackage{makecell,rotating,multirow,diagbox}
\allowdisplaybreaks[4]
\usepackage{algorithm}
\usepackage{algorithmic}
\usepackage{setspace}

\usepackage{multicol}
\usepackage{xcolor}
\usepackage{multirow}
\usepackage{threeparttable}
\usepackage{framed} 
\usepackage{cancel}
\usepackage{booktabs}
\usepackage{tabularx,makecell,multirow}

\newtheorem{theorem}{\textbf{Theorem}}

\newtheorem{lemma}{\textbf{Lemma}}

\newtheorem{corollary}{\textbf{Corollary}}

\newtheorem{proposition}{\textbf{Proposition}}

\makeatletter

\newcommand{\Rmnum}[1]{\expandafter\@slowromancap\romannumeral #1@}
\newcommand{\tabincell}[2]{\begin{tabular}{@{}#1@{}}#2\end{tabular}}

\makeatother

\begin{document}
	\bstctlcite{ref:BSTcontrol}
	
	\title{Multi-Functional Reconfigurable Intelligent Surface: System Modeling and Performance Optimization
	}
	\author{
		Wen~Wang,
		Wanli~Ni,~\IEEEmembership{Member,~IEEE,}
		Hui~Tian,~\IEEEmembership{Senior~Member,~IEEE,} \\
		Yonina~C. Eldar,~\IEEEmembership{Fellow,~IEEE,} 
		and Rui~Zhang,~\IEEEmembership{Fellow,~IEEE}
		\thanks{The work of Hui Tian was supported by the National Key R\&D Program of China under Grant No. 2020YFB1807800. The work of Wen Wang was supported by the Beijing University of Posts and Telecommunications (BUPT) Excellent Ph.D. Students Foundation under Grant CX2022103, and the China Scholarship Council. This work was presented in part at the IEEE International Conference on Acoustics, Speech and Signal Processing (ICASSP), Rhodes Island, Greece, 2023, pp. 1-5, doi: 10.1109/ICASSP49357.2023.10096886. \emph{(Corresponding author: Hui Tian.)}}
		\thanks{W. Wang, W. Ni, and H. Tian are with the State Key Laboratory of Networking and Switching Technology, Beijing University of Posts and Telecommunications, Beijing 100876, China (e-mail:\{wen.wang, charleswall, tianhui\}@bupt.edu.cn).
		 }
		\thanks{Y. C. Eldar is with the Math and CS Faculty, Weizmann Institute of Science, Rehovot 7610001, Israel (e-mail: yonina.eldar@weizmann.ac.il).}
		\thanks{R. Zhang is with School of Science and Engineering, Shenzhen Research Institute of Big Data, The Chinese University of Hong Kong, Shenzhen, Guangdong 518172, China (e-mail: rzhang@cuhk.edu.cn). He is also with the Department of Electrical and Computer Engineering, National University of Singapore, Singapore 117583 (e-mail: elezhang@nus.edu.sg).
    	} 
    	}
	\maketitle
	\begin{abstract}
	    In this paper, we propose and study a multi-functional reconfigurable intelligent surface (MF-RIS) architecture.
		In contrast to conventional single-functional RIS (SF-RIS) that only reflects signals, the proposed MF-RIS simultaneously supports multiple functions with one surface, including reflection, refraction, amplification, and energy harvesting of wireless signals.
		As such, the proposed MF-RIS is capable of significantly enhancing RIS signal coverage by amplifying the signal reflected/refracted by the RIS with the energy harvested.
	    We present the signal model of the proposed MF-RIS, and formulate an optimization problem to maximize the sum-rate of multiple users in an MF-RIS-aided non-orthogonal multiple access network.
		We jointly optimize the transmit beamforming, power allocations as well as the operating modes and parameters for different elements of the MF-RIS and its deployment location, via an efficient iterative algorithm.
       Simulation results are provided which show significant performance gains of the MF-RIS over SF-RISs with only some of its functions available.
       Moreover, we demonstrate that there exists a fundamental trade-off between sum-rate maximization and harvested energy maximization.
       In contrast to SF-RISs which can be deployed near either the transmitter or receiver, the proposed MF-RIS should be deployed closer to the transmitter for maximizing its communication throughput with more energy harvested.
	\end{abstract}
	
	\section{Introduction}
   Reconfigurable intelligent surfaces (RISs) or intelligent reflecting surfaces (IRSs) have emerged as a promising paradigm for future communication networks, due to their merits in improving energy-efficiency and spectrum-efficiency in a low-cost manner\cite{RIS-review-1,RIS_review_Liu}.
	Through modifying the phase shift and/or amplitude of incident signals, RISs are able to establish a tunable communication environment for achieving various objectives, such as throughput maximization \cite{RIS-throughout,User-RIS}, security enhancement \cite{SF-RIS-secure-2,SF-RIS-secure-WW}, energy reduction \cite{SF-RIS-energy consumption-3}, and improved performance fairness \cite{SF-RIS-user fairness-3}.
	However, due to hardware constraints, conventional single-functional RIS (SF-RIS) only supports signal reflection or refraction/transmission, and thus can only serve users located on one side of a RIS\cite{RIS-throughout,User-RIS}.
	The half-space coverage issue faced by SF-RIS greatly restricts the flexibility and effectiveness when deploying SF-RISs in wireless networks with randomly distributed users.

    To overcome this limitation, the authors of \cite{STAR-Liu,IOS-Zhang,TWC-STAR-RIS} proposed the simultaneously transmitting and reflecting reconfigurable intelligent surface (STAR-RIS) and intelligent omni-surface (IOS), by supporting signal reflection and refraction with one surface at the same time.
   Compared to SF-RIS \cite{RIS-throughout,User-RIS,SF-RIS-secure-2,SF-RIS-energy consumption-3,SF-RIS-user fairness-3,SF-RIS-secure-WW}, such dual-functional RIS (DF-RIS) \cite{TWC-STAR-RIS,STAR-Liu,IOS-Zhang} is able to create a ubiquitous smart radio environment by providing full-space signal coverage.
   Moreover, the works \cite{STAR-CL-coverge} and \cite{Wen-DF-RIS} studied the benefits of DF-RIS-aided multi-user communication networks in terms of coverage extension and security enhancement, respectively.
   However, due to the fact that the signal reflected/refracted by the RIS is attenuated twice, the signal path loss can be severe for both SF- and DF-RISs.
      
     With the aim of combating the double attenuation faced by existing passive RISs, recent works such as \cite{active-RIS-TWC-EE} and \cite{active-RIS-analysis-1} proposed an active RIS architecture by embedding power amplifiers into conventional SF-RISs.
     The theoretical and simulation results in \cite{active-RIS-analysis-2} and \cite{active-RIS-WCL-rate} show that by properly designing the phase shift and amplification factors, active RISs yield significant spectrum efficiency gains compared to passive RISs.
     In addition, the authors of \cite{DMAs-WC} proposed dynamic metasurface antennas (DMAs), which enable different levels of amplification and phase shift on incident signals.
     However, both active RISs and DMAs require additional energy consumption to maintain the operation of active components, which makes their performance highly dependent on the availability of external power supply.
    In practice, the aforementioned RIS architectures are powered by battery or the grid\cite{RIS-review-1,RIS_review_Liu}.
    For battery-powered RISs, embedded batteries only provide limited lifetime and cannot support long-term operation.
    Replacing the battery of RISs manually may be costly and impractical\cite{sustainable-RIS-TVT-WPT}.
    The deployment locations of grid-powered RISs are limited because not all places are reachable with power line networks\cite{sustainable-RIS-TCOM,sustainable-RIS-TC-WPT}. 
    Therefore, it is important to develop RIS architectures that are capable of self-sustainability while maintaining the performance advantages of state-of-the-art RISs.
     
     \begin{table*}[t]
     	\caption{Comparison of the proposed MF-RIS with SF- and DF-RIS}
       	\vspace{-2mm}
     	\centering
     	\includegraphics[width=6.6in]{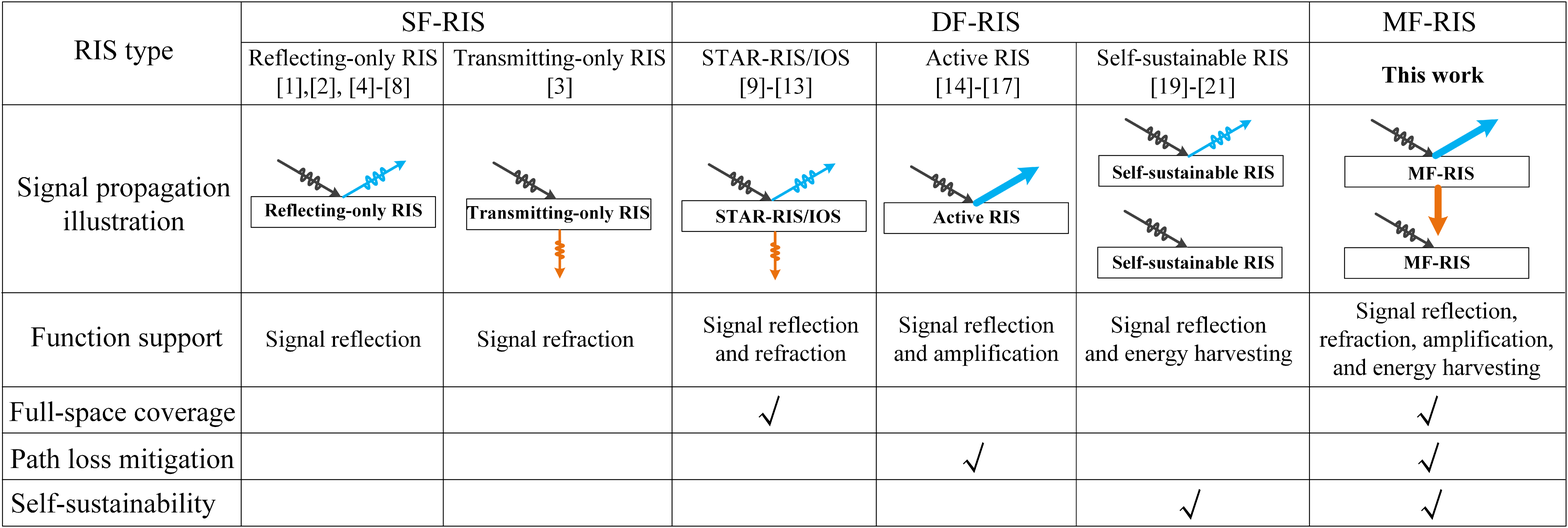}
     	\label{illustration_RIS}
     \end{table*}
 
     In this paper, we propose a new multi-functional RIS (MF-RIS) architecture aiming to overcome the aforementioned drawbacks faced by existing RISs, such as half-space coverage, double attenuation, and reliance on battery/grid.
     Specifically, the proposed MF-RIS utilizes the energy harvested from radio-frequency (RF) signals to support the simultaneous reflection, refraction/transmission, and amplification of incident signals.
     In Table \ref{illustration_RIS}, we compare existing SF- and DF-RIS with the proposed MF-RIS in terms of signal propagation model and design metrics.
     It can be seen that the DF-RISs only partially address the challenges that limit the flexibility and effectiveness of SF-RISs.
     In contrast, the proposed MF-RIS is able to achieve full-space coverage and path loss mitigation in a self-sustainable manner, thereby providing efficient and uninterrupted communication services to users in the whole space.
     By allowing all the elements to flexibly switch between different operating modes, MF-RIS offers more degrees of freedom (DoFs) for signal manipulation. 
     
     To validate the throughput performance improvement when applying MF-RIS in wireless networks, we investigate a sum-rate maximization problem in an MF-RIS-aided non-orthogonal multiple access (NOMA) network. The combination of NOMA and MF-RIS is envisioned as a practically appealing strategy. NOMA enables flexible and efficient resource allocation for MF-RIS assisted multi-user networks by serving multiple users within the same resource block. 
     Meanwhile, MF-RIS facilitates the implementation of NOMA by constructing favorable channels for NOMA.
	The main contributions of this paper are summarized as follows:
	\begin{itemize}
		\item 
		We propose an MF-RIS architecture with multiple functions such as signal reflection, refraction, amplification, and energy harvesting. 
		\item 
		We formulate an optimization problem to maximize the sum-rate of multiple users in an MF-RIS-aided NOMA network by jointly optimizing the transmit beamforming, power allocations as well as the operating modes and parameters for different elements of the MF-RIS and its deployment location.
		This problem is a mixed-integer non-linear programming (MINLP) problem.
		We propose an alternating optimization (AO)-based algorithm to find a  suboptimal solution efficiently.
		\item 
		Extensive simulation results are provided which show that:
		1)
		compared to the conventional passive RIS and self-sustainable RIS, the proposed MF-RIS attains $23.4$\% and $98.8$\% sum-rate gains under the same total power budget, respectively;
		2) due to the limited number of RIS elements, there exists a fundamental trade-off between sum-rate maximization and energy harvesting performance for the MF-RIS;
		3) the proposed MF-RIS should be deployed closer to the transmitter for maximizing its communication throughput with more energy harvested.
	\end{itemize}
	
	The rest of this paper is organized as follows. 
	Section \ref{Preliminary} provides the operation design and signal model of MF-RIS.
	Section \ref{System Model and Problem Formulation} presents the system model and problem formulation of an MF-RIS-aided NOMA network. 
	The resulting MINLP problem is solved in Section \ref{Proposed Solution}.
	Numerical results are presented in Section \ref{Numerical Results}, followed by conclusions in Section \ref{Conclusion}.
	
	\section{Operation Design and Signal Model of MF-RIS}\label{Preliminary}
	In this section, we first introduce the operation mechanism of the proposed MF-RIS.
	Then, we present the signal model of MF-RIS-aided wireless communications.

   	\begin{figure*}[t]
   	\centering
   	\includegraphics[width=6.2 in]{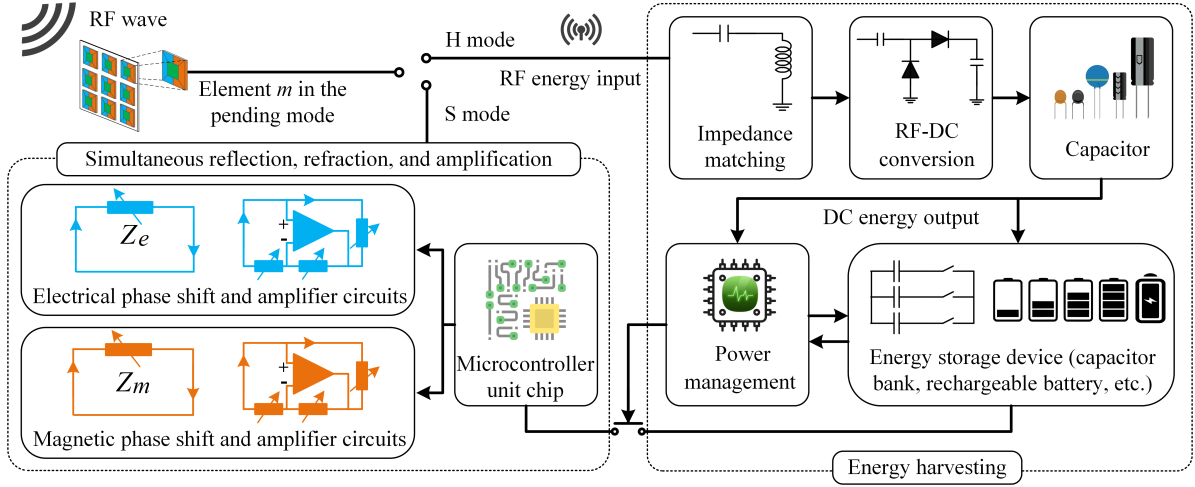}
   	\vspace{-3mm}
   	\caption{A schematic diagram of the proposed MF-RIS.}
   	\label{block diagram_MF-RIS}
   \end{figure*}

	\subsection{Operation Design}	
	As shown in Fig. \ref{block diagram_MF-RIS}, each element of the MF-RIS can operate in two modes: energy harvesting mode (H mode) and signal relay mode (S mode).
	By flexibly adjusting the circuit connection, each element can switch between the H mode and S mode. 
	The elements operating in H mode harvest RF energy from the incident signal, and convert it to direct current (DC) power for supporting the operation of the MF-RIS.
	The energy harvesting circuit contains the following components\cite{2014-RF-EH,2018-rectifier}:
	\begin{itemize}
		\item An impedance matching network consisting of a factor resonator is invoked to guarantee maximum power transmission from the element to the rectifier block.
		\item An RF-DC conversion circuit rectifies the available RF power into DC voltage. 
		\item Capacitors are used to deliver the current to the energy storage device, or as a short-term reserve when RF energy is unavailable.
		\item A power management block decides whether to store the converted electricity energy or use it immediately for signal reflection, refraction, and amplification.
		\item Energy storage devices (e.g., rechargeable batteries and super capacitors) are used to store energy. 
		Whenever the harvested energy exceeds consumption, any excess is stored for future use, thus achieving self-sustainability.
	\end{itemize}

	For other elements operating in S mode, the incident signals are divided into two parts by manipulating the electric and magnetic currents.
	One part is reflected to the reflection half-space and the other is refracted to the other refraction half-space.
	With the aid of a microcontroller unit, these elements leverage the harvested energy to sustain the operation of phase-shifting and amplifier circuits.
	Therefore, the proposed MF-RIS does not need any external power supply in principle.
	The schematic diagrams of the reflection and refraction amplifiers are also shown in Fig. \ref{block diagram_MF-RIS}, where the operational amplifier-based current-inverting converters
	are used to generate the reflected and refracted signals with desired amplification.
	Regarding practical implementation of MF-RIS, there have been many research contributions on the prototype design of signal reflection\cite{Prototype_Passive-RIS}, refraction\cite{STAR-RIS-NTT,IOS-Zeng}, amplification\cite{active-RIS-analysis-1,Active-hardware-1}, and wireless power transfer\cite{WPT,Multi-level,Thin,Power-management}. These existing prototypes can be used to implement the proposed MF-RIS.
	
	\subsection{Signal Model}
	To characterize the signal model of MF-RIS, we consider an MF-RIS with $M$ elements, indexed by $\mathcal{M}=\{1,2,\cdots,M\}$.
	Let	$s_m$ denote the signal received by the $m$-th element.
	Each element cannot simultaneously work in both H and S modes.
    The signals harvested, reflected, and refracted by the $m$-th element are modeled as $y_m^h=(1-\alpha_{m})s_m$, $y_m^r=\alpha_{m}\sqrt{\beta_m^r}e^{j\theta_m^r}s_m$, and $y_m^t=\alpha_{m}\sqrt{\beta_m^t}e^{j\theta_m^t}s_m$, respectively, where $\alpha_{m}\in\{0,1\}$, $\theta_m^r, \theta_m^t\in[0,2\pi)$, and $\beta_{m}^r,\beta_{m}^t\in\left[0,\beta_{\max}\right]$ denote the energy harvesting coefficient, the reflective and refractive phase shifts, and the reflective and refractive amplitude coefficients, respectively.
	Here, $\alpha_m=1$ implies that the $m$-th element operates in S mode, while $\alpha_m=0$ implies that it works in H mode, and $\beta_{\rm max}\geq 1$ denotes the amplification factor.
	The energy consumed by the amplifier should not exceed the maximum available energy that can be applied by the MF-RIS, i.e., $\beta_m^r+\beta_m^t\leq \beta_{\rm max}$.
	The reflective and refractive coefficients are modeled as $\boldsymbol{\Theta}_{r}={\rm{diag}}(\alpha_1\sqrt{\beta_1^r}e^{j\theta_1^r},\alpha_2\sqrt{\beta_2^r}e^{j\theta_2^r},\cdots,\alpha_M\sqrt{\beta_M^r}e^{j\theta_M^r})$ and
	$\boldsymbol{\Theta}_{t}={\rm{diag}}(\alpha_1\sqrt{\beta_1^t}e^{j\theta_1^t},\alpha_2\sqrt{\beta_2^t}e^{j\theta_2^t},\cdots,\alpha_M\sqrt{\beta_M^t}e^{j\theta_M^t})$, respectively, where $\alpha_m\in\{0,1\}$, $\beta_m^r, \beta_m^t\in[0,\beta_{\rm max}]$, $\beta_m^r+\beta_m^t\leq \beta_{\rm max}$, and $\theta_m^r,\theta_m^t\in[0,2\pi)$.
	
	The design variables and constraints for SF-, DF-, and MF-RIS are summarized in Table~\ref{Comparison-variables}.
	We observe that, SF- and DF-RIS can be regarded as special cases of the proposed MF-RIS.
	For example, when $\alpha_{m}\!=\!1$ and $\beta_{\rm max}\!=\!1$, MF-RIS reduces to the STAR-RIS in \cite{STAR-Liu};
    when $\alpha_{m}=1$, $\beta_{\rm max}\!=\!1$, $\beta_{m}^t\!=\!0$, and $\theta_m^t\!=\!0$, MF-RIS reduces to the reflecting-only RIS in \cite{RIS-review-1}.

	 \begin{table*}[t]
	   \caption{Comparison of SF-, DF-, and MF-RIS in terms of design variables}
		\centering
		\label{Comparison-variables}
		\vspace{-6mm}
		\begin{threeparttable}
			\begin{center}
				\renewcommand{\arraystretch}{1.05}
				\scalebox{0.93}{
					\begin{tabular}{|c|l|l|l|l|}
						\hline 
						\multicolumn{2}{|c|}{\multirow{2}{*}{\bf{RIS architecture}}}   & \multicolumn{3}{c|}{\bf{Design variable}}  \\ 
						\cline{3-5}
						\multicolumn{2}{|c|}{} & Energy harvesting  & Amplitude coefficient &  Phase shift     \\   
						\hline 
						\multirow{2}{*}{SF-RIS}  & Reflecting-only RIS\cite{RIS-review-1}  & $\alpha_{m}=1$ &  $\beta_m^r\in[0,1]$ &  $\theta_m^r\in[0,2\pi)$ \\
						\cline{2-5} 
						& Transmitting-only RIS\cite{User-RIS
						} & $\alpha_{m}=1$   &  $\beta_m^t\in[0,1]$ &  $\theta_m^t\in[0,2\pi)$    \\
						\hline 
						\multirow{3}{*}{DF-RIS}  & STAR-RIS/IOS\cite{STAR-Liu} & $\alpha_{m}=1$   &  $\beta_m^r,\beta_m^t\in[0,1]$ &  $\theta_m^r,\theta_m^t\in[0,2\pi)$ \\
						\cline{2-5} 
						& Active RIS \cite{active-RIS-TWC-EE} &  $\alpha_{m}=1$   &  $\beta_m^r\in[0,\beta_{\rm max}]$ &  $\theta_m^r\in[0,2\pi)$    \\		
						\cline{2-5}
						& Self-sustainable RIS\cite{sustainable-RIS-TVT-WPT} & $\alpha_{m}\in\{0,1\}$ &  $\beta_m^r\in[0,1]$ &  $\theta_m^r\in[0,2\pi)$    \\		
						\hline
						\multicolumn{2}{|c|}{\bf{MF-RIS (This work)}}  &  $\alpha_{m}\in\{0,1\}$ & $\beta_m^r,\beta_m^t\in[0,\beta_{\rm max}]$, $\beta_m^r+\beta_m^t\leq \beta_{\rm max}$ & $\theta_m^r,\theta_m^t\in[0,2\pi)$  \\
						\hline 		
					\end{tabular}
				}
			\end{center}
		\end{threeparttable} 
	\end{table*}
    
	\section{System Model and Problem Formulation}\label{System Model and Problem Formulation}
	\subsection{Network Model}	
	We consider an MF-RIS assisted downlink NOMA network, where an $N$-antenna BS serves $J$ single-antenna users with the aid of an MF-RIS consisting of $M$ elements.
	We denote the reflection (refraction) spatial direction as $r$ ($t$). 
	The spatial direction set and the user set are denoted by $\mathcal{K}\!=\!\{r,t\}$ and $\mathcal{J}\!=\!\{1,2,\cdots,J\}$, respectively.
	We denote $\mathcal{J}_k\!=\!\{1,2,\cdots,J_k\}$ as the set of users located in direction $k$ and $\mathcal{J}_r\cup\mathcal{J}_t\!=\!\mathcal{J}$.
	For notation simplicity, we index user $j$ in direction $k$ by $U_{kj}$.
	Furthermore, we consider a three-dimensional (3D) Cartesian coordinate system with the locations of the BS, MF-RIS, and user $U_{kj}$ being $\mathbf{w}_b\!=\![x_b,y_b,z_b]^{\rm T}$, $\mathbf{w}\!=\![x,y,z]^{\rm T}$, and $\mathbf{w}_{kj}\!=\![x_{kj},y_{kj},0]^{\rm T}$, respectively.
	In practice, due to the limited coverage of MF-RIS, its deployable region is also limited.
	Denote by $\mathcal{P}$ the predefined region for MF-RIS deployment. 
	Then the following constraint should be satisfied:
	\setlength{\abovedisplayskip}{3pt}
	\setlength{\belowdisplayskip}{3pt}
	\begin{eqnarray}
	    \nonumber
		\mathbf{w}\in\mathcal{P} &\!\!\!\!\!=\big\{  \left[x,y,z\right]^{\rm T}|x_{\rm min}\leq x\leq x_{\rm max}, y_{\rm min}\leq y\leq y_{\rm max}, \\
		\label{MF-RIS region}
		&\!\!\!\!\!\!\!\!\!\!\!\! z_{\rm min}\leq z\leq z_{\rm max}\big\},
	\end{eqnarray}
	where $\left[x_{\rm min},x_{\rm max}\right]$, $\left[y_{\rm min},y_{\rm max}\right]$, and $\left[z_{\rm min},z_{\rm max}\right]$ denote the candidate ranges along the $X$-, $Y$- and $Z$-axes, respectively.
	
	To characterize the performance upper bound that can be achieved by MF-RIS, we assume that perfect channel state information of all channels is available.
	Since the BS and the MF-RIS are usually deployed at relatively high locations, the line-of-sight (LoS) links can be exploited for them.
	Therefore, similar to existing RIS works\cite{SF-RIS-energy consumption-3,SF-RIS-user fairness-3}, we adopt Rician fading for all channels.
	For instance, the channel matrix $\mathbf{H}\in\mathbb{C}^{M\times N}$	between the BS and the MF-RIS is given by
	\begin{align}
		\label{H_bs}
		\mathbf{H}=\underbrace{\sqrt{h_0d_{bs}^{-\kappa_{0}}}}_{L_{bs}} \underbrace{\left(\sqrt{\frac{\beta_{0}}{\beta_{0}+1}}\mathbf{H}^{\rm{LoS}}+\sqrt{\frac{1}{\beta_{0}+1}}\mathbf{H}^{\rm{NLoS}}\right)}_{\hat{\mathbf{H}}},
	\end{align}
	where $L_{bs}$ is the distance-dependent path loss, and $\hat{\mathbf{H}}$ is composed of the array response and small-scale fading.
	Specifically, $h_0$ is the path loss at the reference distance of 1 meter (m), $d_{bs}$ is the link distance between the BS and the MF-RIS, and $\kappa_{0}$ is the corresponding path loss exponent. 
	As for the small-scale fading, $\beta_{0}$ is the Rician factor, and $\mathbf{H}^{\rm NLoS}$ is the non-LoS component that follows independent and identically distributed (i.i.d.) Rayleigh fading.
	Assuming that the MF-RIS is placed parallel to the $Y\!-\!Z$ plane and its $M$ elements form an $M_y\times M_z\!=\!M$ uniform rectangular array, the LoS component $\mathbf{H}^{\rm{LoS}}$ is expressed as\cite{2005-AOD-AOA}
	\begin{align}
	 \nonumber
     \mathbf{H}^{\rm{LoS}}&\!\!=\!\!\big[1,e^{-j\frac{2\pi}{\lambda}d\sin\varphi_{r}\sin\vartheta_{r}},\cdots,e^{-j\frac{2\pi}{\lambda}(M_z-1)d\sin\varphi_{r}\sin\vartheta_{r}}\big]^{\rm T}\\
    \nonumber
    &\!\! \otimes \!\!\big[1,e^{-j\frac{2\pi}{\lambda}d\sin\varphi_{r}\cos\vartheta_{r}},\cdots,e^{-j\frac{2\pi}{\lambda}(M_y-1)d\sin\varphi_{r}\cos\vartheta_{r}}\big]^{\rm T}\\
    \label{H_bs_los}
    &\!\! \otimes\!\! \big[1,e^{-j\frac{2\pi}{\lambda}d\sin\varphi_{t}\cos\vartheta_{t}},\cdots,e^{-j\frac{2\pi}{\lambda}(N-1)d\sin\varphi_{t}\cos\vartheta_{t}}\big],
	\end{align}
	where $\otimes$ denotes the Kronecker product, $\lambda$ is the carrier wavelength, and $d$ is the antenna separation.
	Here, $\varphi_{r}$, $\vartheta_{r}$, $\varphi_{t}$, and $\vartheta_{t}$ represent the vertical and horizontal angle-of-arrivals, and the vertical and horizontal angle-of-departures of this LoS link, respectively.
	
	The channel vectors from the BS to user $U_{kj}$ and from the MF-RIS to user $U_{kj}$, denoted by $\mathbf{h}_{kj}^{\mathrm H}\in\mathbb{C}^{1\times N}$ and $\mathbf{g}_{kj}^{\mathrm H}\in\mathbb{C}^{1\times M}$, are generated by a process similar to obtaining $\mathbf{H}$, and are given by
	\begin{subequations}
		\begin{align}
			\label{h_bkj}
			\!\!\!\mathbf{h}_{kj}&=\underbrace{\sqrt{h_0d_{bkj}^{-\kappa_{1}}}}_{L_{bkj}} \underbrace{\left(\sqrt{\frac{\beta_{1}}{\beta_{1}+1}}\mathbf{h}_{kj}^{\rm{LoS}}+\sqrt{\frac{1}{\beta_{1}+1}}\mathbf{h}_{kj}^{\rm{NLoS}}\right)}_{\hat{\mathbf{h}}_{kj}}, \\
			\label{h_skj}
			\!\!\!\mathbf{g}_{kj}&=\underbrace{\sqrt{h_0d_{skj}^{-\kappa_{2}}}}_{L_{skj}} \underbrace{\left(\sqrt{\frac{\beta_{2}}{\beta_{2}+1}}\mathbf{g}_{kj}^{\rm{LoS}}+\sqrt{\frac{1}{\beta_{2}+1}}\mathbf{g}_{kj}^{\rm{NLoS}}\right)}_{\hat{\mathbf{g}}_{kj}}.
		\end{align}
	\end{subequations}

	\subsection{MF-RIS-Aided Downlink NOMA}
	Using the NOMA protocol, the BS transmits the superimposed signal by
	exploiting multiple beamforming vectors, i.e., $\mathbf{s}=\sum_{k\in\mathcal{K}}\mathbf{f}_k\sum_{j\in\mathcal{J}_k}\sqrt{p_{kj}}s_{kj}$.
	Here, $\mathbf{f}_k$ is the transmit beamforming vector for direction $k$, satisfying $\sum\nolimits_{k\in\mathcal{K}} \lVert \mathbf{f}_k\lVert^2 \leq P_{\rm BS}^{\max}$, where $P_{\rm BS}^{\max}$ denotes the maximum transmit power of the BS. 
	Moreover, $p_{kj}$ is the power allocation factor of user $U_{kj}$ with $\sum_{j\in\mathcal{J}_k} p_{kj}=1$, and $s_{kj}\!\sim\! \mathcal{CN}(0,1)$ denotes the modulated data symbol, which is independent over $k$.
	Therefore, the signal received at user $U_{kj}$ is given by\cite{STAR-NOMA-Zuo}
	\begin{align}
		\nonumber
		y_{kj}=&\underbrace{\bar{\mathbf{h}}_{kj}\mathbf{f}_{k}\sqrt{p_{kj}}s_{kj}}_{\rm desired \ signal} +\underbrace{\bar{\mathbf{h}}_{kj}\mathbf{f}_{k}\sum\nolimits_{i\in\{\mathcal{J}_k/j\}}\sqrt{p_{ki}}s_{ki}}_{\rm intra-space \ interference} \\
		\label{received_signal}
		& +  \underbrace{\bar{\mathbf{h}}_{kj}\mathbf{f}_{\bar{k}}\sum\nolimits_{i\in\mathcal{J}_{\bar{k}}}\sqrt{p_{\bar{k}i}}s_{\bar{k}i}}_{\rm inter-space \ interference} 
		+\underbrace{\mathbf{g}_{kj}^{\mathrm H}\boldsymbol{\Theta}_{k}\mathbf{n}_s}_{\rm RIS\  noise}+ n_{kj}, 
	\end{align}
	where $\bar{k}=r$, if $k=t$; and $\bar{k}=t$, if $k=r$, $\mathbf{n}_s\sim \mathcal{C} \mathcal{N}(\mathbf{0}, \sigma_s^{2}\mathbf{I}_M)$ denotes the amplification noise introduced at the MF-RIS with per-element noise power $\sigma_s^2$, and $n_{kj} \sim \mathcal{C} \mathcal{N}(0, \sigma_u^{2})$ denotes additive white Gaussian noise (AWGN) at user $U_{kj}$ with power $\sigma_u^2$.
	In addition, $\bar{\mathbf{h}}_{kj}=\mathbf{h}_{kj}^{\rm H}+\mathbf{g}_{kj}^{\rm H}\boldsymbol{\Theta}_k\mathbf{H}$ represents the combined channel vector from the BS to user $U_{kj}$.
	For conventional passive RISs, the term $\mathbf{g}_{kj}^{\mathrm H}\boldsymbol{\Theta}_{k}\mathbf{n}_s$ is negligibly small compared to the AWGN at user $U_{kj}$ and thus usually omitted. 
	However, such noise is amplified by the amplification unit in our considered MF-RIS and thus can no longer be ignored.
	
	Following the NOMA protocol, all users employ successive interference cancellation (SIC) to detect the signal and remove interference\cite{RIS-throughout}.
    We assume that the users in direction $k$ are ranked in an ascending order according to the equivalent combined channel gains, expressed as
	\begin{align}
         \nonumber
		&\frac{|\bar{\mathbf{h}}_{kj}\mathbf{f}_{k}|^2p_{kj}}
		{|\bar{\mathbf{h}}_{kj}\mathbf{f}_{k}|^2P_{kj}+|\bar{\mathbf{h}}_{kj}\mathbf{f}_{\bar{k}}|^2+\sigma_s^2\lVert\mathbf{g}_{kj}^{\mathrm H}\boldsymbol{\Theta}_{k}\lVert^2+\sigma_u^2} \\
		\label{original_decoding_order}
		& \leq 	\frac{|\bar{\mathbf{h}}_{kl}\mathbf{f}_{k}|^2p_{kj}}
		{|\bar{\mathbf{h}}_{kl}\mathbf{f}_{k}|^2P_{kj}+|\bar{\mathbf{h}}_{kl}\mathbf{f}_{\bar{k}}|^2+\sigma_s^2\lVert\mathbf{g}_{kl}^{\mathrm H}\boldsymbol{\Theta}_{k}\lVert^2+\sigma_u^2}, 
	\end{align}
   where $k\in\mathcal{K}$, $j\in\mathcal{J}_k$, $l\in\mathcal{L}_k\!=\!\{j,j+1,\cdots,J_k\}$, and $P_{kj}\!=\!\sum_{i=j+1}^{J_k}p_{ki}$. 
   The SIC condition in (\ref{original_decoding_order}) can be equivalently transformed into
   the following inequality:
     \begin{align}
     	\nonumber
    	& \!\!\!\! |\bar{\mathbf{h}}_{kj}\mathbf{f}_{k}|^2p_{kj} (|\bar{\mathbf{h}}_{kl}\mathbf{f}_{k}|^2P_{kj}+|\bar{\mathbf{h}}_{kl}\mathbf{f}_{\bar{k}}|^2+\sigma_s^2\lVert\mathbf{g}_{kl}^{\mathrm H}\boldsymbol{\Theta}_{k}\lVert^2+\sigma_u^2) \\
    	\label{decoding_order_1}
    	& \!\!\!\!\leq \! |\bar{\mathbf{h}}_{kl}\mathbf{f}_{k}|^2p_{kj} (|\bar{\mathbf{h}}_{kj}\mathbf{f}_{k}|^2P_{kj} \!+ \! |\bar{\mathbf{h}}_{kj}\mathbf{f}_{\bar{k}}|^2 \! + \! \sigma_s^2\lVert\mathbf{g}_{kj}^{\mathrm H}\boldsymbol{\Theta}_{k}\lVert^2 \! + \!\sigma_u^2).  \!\!\!\!
    \end{align}
   By subtracting the term $p_{kj}|\bar{\mathbf{h}}_{kj}\mathbf{f}_{k}|^2|\bar{\mathbf{h}}_{kl}\mathbf{f}_{k}|^2P_{kj}$ from both sides of (\ref{decoding_order_1}) and dividing by $p_{kj}$, we obtain the following inequality for all $k,j,l$:
    \begin{align}
    	\label{decoding_order}
    	\frac{|\bar{\mathbf{h}}_{kj}\mathbf{f}_{k}|^2}
    	{|\bar{\mathbf{h}}_{kj}\mathbf{f}_{\bar{k}}|^2+\sigma_s^2\lVert\mathbf{g}_{kj}^{\mathrm H}\boldsymbol{\Theta}_{k}\lVert^2+\sigma_u^2}   \leq 	\frac{|\bar{\mathbf{h}}_{kl}\mathbf{f}_{k}|^2}
    	{|\bar{\mathbf{h}}_{kl}\mathbf{f}_{\bar{k}}|^2+\sigma_s^2\lVert\mathbf{g}_{kl}^{\mathrm H}\boldsymbol{\Theta}_{k}\lVert^2+\sigma_u^2}.
    \end{align}
    We observe from (\ref{decoding_order}) that the SIC condition is independent of the power allocation coefficients $\{p_{kj}\}$.
    
	Based on (\ref{decoding_order}), for any users $U_{kj}$ and $U_{kl}$ satisfying $j\leq l$, the achievable rate for user $U_{kl}$ to decode the intended signal of user $U_{kj}$ is given by
	\begin{align}
		\label{R_kj}
		R_{l\to j}^k \!\!  = \! \! \log_2\left(1 \! + \!\frac{|\bar{\mathbf{h}}_{kl}\mathbf{f}_{k}|^2p_{kj}}
		{|\bar{\mathbf{h}}_{kl}\mathbf{f}_{k}|^2P_{kj} \! + \! |\bar{\mathbf{h}}_{kl}\mathbf{f}_{\bar{k}}|^2 \!  +\! \sigma_s^2\lVert\mathbf{g}_{kl}^{\mathrm H}\boldsymbol{\Theta}_{k}\lVert^2+\sigma_u^2}\right).
	\end{align}
   To guarantee the success of SIC, the achievable signal-to-interference-plus-noise ratio (SINR) at user $U_{kl}$ to
	decode the signal of user $U_{kj}$ for all $j\leq l$ should be no less than the SINR at user $U_{kj}$ to decode its own signal.
	Thus, we have the following SIC decoding rate constraint:
	\begin{align}
		\label{rate-SIC}
		R_{l\to j}^k \geq R_{j\to j}^k , ~\forall k \in\mathcal{K}, \forall j\in\mathcal{J}_k, \forall l\in\mathcal{L}_k.
	\end{align}
    
	\subsection{Power Dissipation Model}
	Define the energy harvesting coefficient matrix of the $m$-th element as 
	\begin{align}
		\mathbf{T}_m={\rm diag}([\underbrace{0,\cdots,}_{1 \ \text{to} \ m-1} 1-\alpha_{m} \underbrace{,\cdots, 0}_{m+1 \ \text{to} \ M}]).
	\end{align}
	Then, the RF power received at the $m$-th element is given by
	\begin{align}
		\label{MF-RIS-RF}
		P_m^{\rm{RF}}=\mathbb{E}\left(\left\|\mathbf{T}_m\left(\mathbf{H}\mathbf{s}+\mathbf{n}_{s}\right)\right\|^{2}\right), 
	\end{align}
	where the expectation operator $\mathbb{E}(\cdot)$ is over $\mathbf{s}$ and $\mathbf{n}_s$.
	
	In order to capture the dynamics of the RF energy conversion efficiency for different input power levels, a non-linear energy harvesting model is adopted\cite{EH-parameter}. Accordingly, the total power harvested at the $m$-th element is given by
	\begin{eqnarray}
		\label{C_energy_coefficients}
		P_m^{\mathrm{A}} \! = \!\frac{\Upsilon_m  \! - \! Z\Omega}{1  \! -  \! \Omega},~\Upsilon_m  \! =  \! \frac{Z}{1+e^{-a (P_m^{\rm RF}-q)}}, ~\Omega  \! =  \! \frac{1}{1+e^{aq}},
	\end{eqnarray}
	where $\Upsilon_m$ is a logistic function with respect to (w.r.t.) the received RF power $P_m^{\rm RF}$, and $Z\geq 0$ is a constant determining the maximum harvested power.
	Constant $\Omega$ is introduced to ensure a zero-input/zero-output response for H mode, with constants $a>0$ and $q>0$ capturing the joint effects of circuit sensitivity limitations and current leakage.
	
	To ensure energy self-sustainability, the power consumed by the proposed MF-RIS should not exceed the power harvested. The power consumption of the proposed MF-RIS is mainly caused by the operation of phase shifters, amplifiers, power conversion circuits, and the output power.
	Other sources of power consumption, such as powering the impedance matching and mode switching circuits, are negligible in comparison\cite{sustainable-RIS-TC-WPT,sustainable-RIS-TCOM,sustainable-RIS-TVT-WPT}.
	Given that the MF-RIS has $2\sum\nolimits_{m\in\mathcal{M}} \alpha_{m}$ phase shifters, $2\sum\nolimits_{m\in\mathcal{M}} \alpha_{m}$ amplifiers, and $M-\sum\nolimits_{m\in\mathcal{M}} \alpha_{m}$ power conversion circuits in operation, we have the following energy self-sustainability constraint:
	\begin{align}
		\nonumber
		& 2\left(P_{b}+P_{\rm DC}\right)\sum\nolimits_{m\in\mathcal{M}} \alpha_{m}+(M-\sum\nolimits_{m\in\mathcal{M}} \alpha_{m})P_{\rm C} \\
		\label{C_energy}
		&+\xi P_{\rm O} \leq \sum\nolimits_{m\in\mathcal{M}} P_m^{\mathrm{A}},
	\end{align}
	where $P_{b}$, $P_{\rm DC}$, and $P_{\rm C}$ denote the power consumed by each phase shifter, the DC biasing power consumed by the amplifier, and the power consumed by the RF-to-DC power conversion circuit, respectively.
	Here, $\xi$ is the inverse of the amplifier efficiency, and $P_{\rm O}=\sum_{k\in\mathcal{K}} \big(\sum_{k'\in\mathcal{K}} \lVert \boldsymbol{\Theta}_{k} \mathbf{H} \mathbf{f}_{k'}\lVert^2$ $+\sigma_s^2 \lVert \boldsymbol{\Theta}_{k} \lVert_F^2 \big)$ represents the output power of the MF-RIS. 
	 
	\subsection{Problem Formulation}
	Our goal is to maximize the sum-rate of all users while maintaining self-sustainability of the MF-RIS by jointly optimizing the power allocation, transmit beamforming at the BS, and the coefficient matrix and 3D location of the MF-RIS.
	The optimization problem is formulated as
	\begin{subequations}
		\label{P0}
		\begin{eqnarray}
			\label{P0_function}
			& \underset{p_{kj},\mathbf{f}_k,\boldsymbol{\Theta}_k,\mathbf{w}}{\max}  & \sum\nolimits_{k\in\mathcal{K}}\sum\nolimits_{j\in\mathcal{J}_k}R_{j\to j}^k\\
			\label{C_power allocation}
			& \operatorname{s.t.} &	\sum\nolimits_{j\in\mathcal{J}_k} p_{kj}=1, ~\forall k \in\mathcal{K},\\
			\label{C_transmit beamforming}
			&&\sum\nolimits_{k\in\mathcal{K}} \lVert \mathbf{f}_k\lVert^2\leq P_{\rm BS}^{\max},\\
			\label{C_Rmin}
			&& R_{j \to j}^k \geq R_{kj}^{\rm min}, ~\forall k\in\mathcal{K}, \forall j\in\mathcal{J}_k,\\ 
			\label{C-MF-RIS}
			&&\boldsymbol{\Theta}_{k}\in \mathcal{R}_{\rm MF}, ~\forall k \in\mathcal{K},\\
			\label{C-HS}
			&&{\rm (\ref{MF-RIS region}),  (\ref{decoding_order}), (\ref{rate-SIC}), (\ref{C_energy})},
		\end{eqnarray}
	\end{subequations}
	where $\mathcal{R}_{\rm MF}$ is the feasible set for MF-RIS coefficients, with $\mathcal{R}_{\rm MF}\!=\!\{\alpha_m,\beta_m^k,\theta_m^k|\alpha_m\in\{0,1\},\beta_m^k\in[0,\beta_{\max}], \sum\nolimits_{k\in\mathcal{K}}\beta_m^k$ $\leq\beta_{\rm max},\theta_m^k\in[0,2\pi),\forall m, k\}$.
		Constraint (\ref{C_power allocation}) represents the power allocation restriction, (\ref{C_transmit beamforming}) ensures that the total transmit power at the BS cannot exceed the power budget $P_{\rm BS}^{\max}$, and (\ref{C_Rmin}) guarantees that the achievable data rate of user $U_{kj}$ is above the quality-of-service (QoS) requirement $R_{kj}^{\rm min}$.
	Constraints (\ref{C-MF-RIS}) and (\ref{MF-RIS region}) specify the allowable ranges of MF-RIS coefficients and locations, respectively, and (\ref{decoding_order}) determines the SIC decoding order of NOMA users. 
	In addition, constraint (\ref{rate-SIC}) ensures successful SIC decoding, and (\ref{C_energy}) guarantees the energy self-sustainability of the MF-RIS.

	\begin{figure*}[t]
		\centering
		\includegraphics[width=6in]{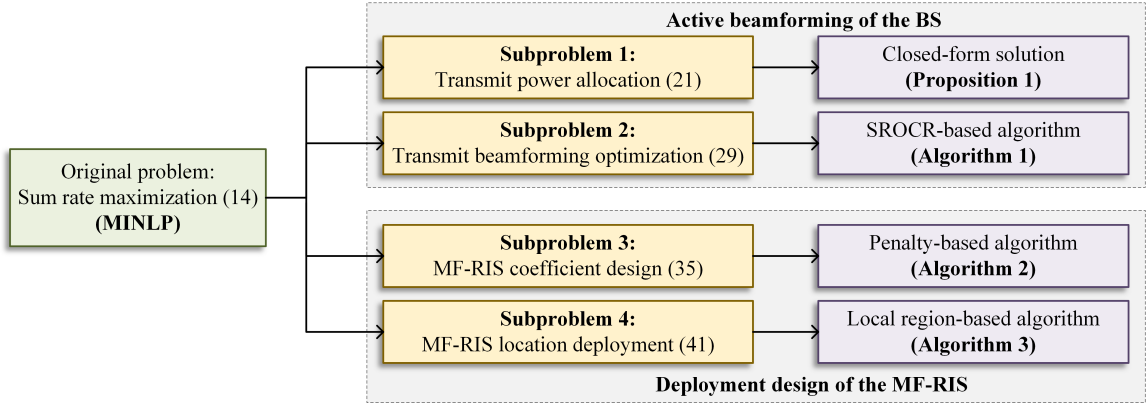}
		\caption{A roadmap for problem decomposition and algorithm design.}
		\label{roadmap}
	\end{figure*} 
	
	The sum-rate maximization problems studied in existing works on self-sustainable RIS\cite{sustainable-RIS-TC-WPT} and STAR-RIS\cite{STAR-NOMA-Zuo} can be regarded as special cases of problem (\ref{P0}).
   However, their results are not applicable to solving (\ref{P0}) due to the following new challenges introduced by MF-RIS:
	1) the objective function (\ref{P0_function}) and constraints (\ref{decoding_order}), (\ref{rate-SIC}), (\ref{C_energy}), and (\ref{C_Rmin}) involve closely coupled variables;
	2) the adopted non-linear energy harvesting model makes constraint (\ref{C_energy}) intractable, compared to the linear model in \cite{sustainable-RIS-TC-WPT};
	3) due to the signal amplification, additional RIS noise needs to be considered in the objective function (\ref{P0_function}) and constraints (\ref{decoding_order}), (\ref{rate-SIC}), (\ref{C_energy}), and (\ref{C_Rmin}), which complicates the resource allocation problem;
	4) the binary energy harvesting coefficients result in combinatorial constraints (\ref{decoding_order}), (\ref{rate-SIC}), (\ref{C_energy}), (\ref{C_Rmin}), and (\ref{C-MF-RIS}), which makes (\ref{P0}) an MINLP problem.
	Therefore, the formulated problem (\ref{P0}) for the MF-RIS is more challenging to solve as compared to those for existing RISs in \cite{sustainable-RIS-TC-WPT} and \cite{STAR-NOMA-Zuo}.
	
	\section{Proposed Solution for Active Beamforming and MF-RIS Deployment} \label{Proposed Solution}
	To solve (\ref{P0}) efficiently, we propose an AO-based algorithm. 	
	As shown in Fig. \ref{roadmap},	the original problem (\ref{P0}) is decomposed into four subproblems.
	Specifically, the power allocation strategy is obtained in closed form based on successive cancellation.
	The transmit beamforming optimization is then solved using the sequential rank-one constraint relaxation (SROCR) method.
	Next, the MF-RIS coefficient is designed by applying the penalty function.
	Finally, the MF-RIS location is determined by adopting local region optimization.
  
   \subsection{Problem Transformation}
   Before solving (\ref{P0}), we transform it into a more tractable form.
   First, we observe that constraint (\ref{rate-SIC}) is a necessary condition for inequality (\ref{decoding_order}), as (\ref{decoding_order}) is equivalent to (\ref{original_decoding_order}).
   This observation shows that under the proposed decoding order, the SIC
   condition is guaranteed, which is consistent with the conclusions obtained in existing NOMA works such as \cite{STAR-NOMA-Zuo} and \cite{Power-TWC}.
    As a result, removing constraint (\ref{rate-SIC}) does not affect the optimality of (\ref{P0}) when (\ref{decoding_order}) is satisfied.
    Hence, (\ref{rate-SIC}) is removed from (\ref{P0}) in the following.
   
   Next, to deal with the non-convex constraint (\ref{decoding_order}), we introduce slack variables as
   $A_{kj}=|\bar{\mathbf{h}}_{kj}\mathbf{f}_{k}|^{-2}$, $B_{kj}=|\bar{\mathbf{h}}_{kj}\mathbf{f}_{\bar{k}}|^2+\sigma_s^2\lVert\mathbf{g}_{kj}^{\mathrm H}\boldsymbol{\Theta}_{k}\lVert^2+\sigma_u^2$, and $\Gamma_{kj}=A_{kj}^{-1}B_{kj}^{-1}$.
   Then, (\ref{decoding_order}) is equivalently transformed into 
   \begin{subequations}
   	\begin{align}
   		\label{P_trans_AB_1}
   		&\!\!\!\!\!\!A_{kj}^{-1}\leq|\bar{\mathbf{h}}_{kj}\mathbf{f}_{k}|^2, ~\forall k\in\mathcal{K},\forall j\in\mathcal{J}_k,\!\!\!\!\!\! \\
   		\label{P_trans_AB_2}
   		&\!\!\!\!\!\!B_{kj}\geq |\bar{\mathbf{h}}_{kj}\mathbf{f}_{\bar{k}}|^2+\sigma_s^2\lVert\mathbf{g}_{kj}^{\mathrm H}\boldsymbol{\Theta}_{k}\lVert^2+\sigma_u^2, ~\forall k\in\mathcal{K},\forall j\in\mathcal{J}_k,\!\!\!\!\!\!\\
   		\label{P_trans_AB_3}
   		&\!\!\!\!\!\!\Gamma_{kj} \geq A_{kj}^{-1}B_{kj}^{-1}, 
   		~\forall k\in\mathcal{K},\forall j\in\mathcal{J}_k,\!\!\!\!\!\!\\
   		\label{P_trans_Gamma-Energy}
   		&\!\!\!\!\!\!\Gamma_{kj} \leq A_{kl}^{-1}B_{kl}^{-1},~\forall k  \in  \mathcal{K},\forall j  \in  \mathcal{J}_k,\forall l\in\mathcal{L}_k.\!\!\!\!\!\!
   	\end{align}
   \end{subequations}
   
   As for the energy self-sustainability constraint (\ref{C_energy}), it is difficult to directly observe and handle due to the non-linear energy harvesting model based on the logistic function. 
   Therefore, we first substitute the terms in (\ref{C_energy_coefficients}) into (\ref{C_energy}), and equivalently rewrite (\ref{C_energy}) in the following form:
   \begin{align}
   	\nonumber
   	& (\mathcal{W}+\xi  P_{\rm O})(1-\Omega)Z^{-1}  +  M\Omega \\
   	\label{R_C_energy_1}
   	&\leq \sum\nolimits_{m\in\mathcal{M}} \big( 1+ \exp(-a(P_m^{\rm{RF}}-q))\big)^{-1},
   \end{align}
   where $\mathcal{W}\!=\!2(P_b+P_{\rm DC})\sum\nolimits_{m\in\mathcal{M}}\alpha_m\!+\!(M-\sum\nolimits_{m\in\mathcal{M}}\alpha_m )P_{\rm C}$.
   The complex right-hand-side (RHS) of (\ref{R_C_energy_1}) and the non-convex expression in (\ref{MF-RIS-RF}) make (\ref{R_C_energy_1}) difficult to deal with.
   By introducing slack variables $\zeta_m\!=\! P_m^{\rm{RF}}$ and $\mathcal{C}_m\!=\!1+ \exp(-a(\zeta_m-q))$, we further recast (\ref{R_C_energy_1}) as
    \begin{subequations}
   \begin{align}
   	\label{R_C_energy_2}
   	&(\mathcal{W}+\xi  P_{\rm O})(1-\Omega)Z^{-1} \! + \! M\Omega\leq \sum\nolimits_{m\in\mathcal{M}} \mathcal{C}_m^{-1}, \\
   	\label{P_trans_C-Energy-2}
   	&\zeta_m \leq P_m^{\rm{RF}},~~\mathcal{C}_m\geq 1+ \exp\big(-a(\zeta_m-q)\big),~\forall m.
   \end{align}
  \end{subequations}

    Constraints (\ref{P_trans_Gamma-Energy}) and (\ref{R_C_energy_2}) are non-convex due to their RHSs.
    Here, we exploit the successive convex approximation (SCA) technique to tackle them.
    Specifically, the lower bounds of their RHSs at the feasible point $\{A_{kl}^{(\ell)},B_{kl}^{(\ell)},\mathcal{C}_m^{(\ell)}\}$ in the $\ell$-th iteration are, respectively, given by
     \begin{subequations}
    \begin{align}
    	\label{P_trans_lb_1}
    	&\Gamma_{kl}^{\rm lb}=\frac{1}{A_{kl}^{(\ell)}B_{kl}^{(\ell)}}-\frac{A_{kl}-A_{kl}^{(\ell)}}{(A_{kl}^{(\ell)})^2B_{kl}^{(\ell)}}
    	-\frac{B_{kl}-B_{kl}^{(\ell)}}{(B_{kl}^{(\ell)})^2A_{kl}^{(\ell)}}, \\
    	\label{P_trans_lb_2}
    	&\mathcal{C}^{\rm lb}=\sum\nolimits_{m\in\mathcal{M}}\left(\frac{2}{\mathcal{C}_m^{(\ell)}}-\frac{\mathcal{C}_m}{(\mathcal{C}_m^{(\ell)})^{2}}\right).
    \end{align}
 \end{subequations}
    As a result, by defining $\bar{W}=\frac{(\mathcal{C}^{\rm lb}-M\Omega)Z}{(1-\Omega)\xi}-\frac{\mathcal{W}}{\xi}$ and $\Delta_0=\{A_{kj}, B_{kj}, \Gamma_{kj}, \mathcal{C}_m, \zeta_m\}$, problem (\ref{P0}) is equivalently transformed into the following one:
  	\begin{subequations}
  	\label{P1}
  	\begin{align}
  		\label{P1_function}
  		 \underset{p_{kj},\mathbf{f}_k,\boldsymbol{\Theta}_k,\mathbf{w},\Delta_0}{\max}  & \sum\nolimits_k\sum\nolimits_{j\in\mathcal{J}_k}R_{j\to j}^k\\
  		\label{C_Gamma_lb}
  		 \operatorname{s.t.}~~~ &
  		\Gamma_{kj} \leq \Gamma_{kl}^{\rm lb}, ~\forall k\in\mathcal{K},\forall j\in\mathcal{J}_k,\forall l\in\mathcal{L}_k, \\
  		& \bar{W} \geq  P_{\rm O},\\
  		&{\rm (\ref{MF-RIS region}),(\ref{C_power allocation}){\text -}(\ref{C-MF-RIS}), (\ref{P_trans_AB_1}){\text -}(\ref{P_trans_AB_3}), (\ref{P_trans_C-Energy-2})}.
  	\end{align}
  \end{subequations}
  
	\subsection{Power Allocation}
	To start with, we focus on the optimization of $\{p_{kj}\}$ with given $\{\mathbf{f}_k,\boldsymbol{\Theta}_k,\mathbf{w}\}$.
	Since the inter-cluster interference is independent of $\{p_{kj}\}$, the power allocation problem can be decomposed into two subproblems\cite{Power-TWC}.
	For direction $k$, the optimization problem is formulated as
	\begin{subequations}
		\label{P_p-1}
		\begin{eqnarray}
			& \underset{p_{kj}}{\max}  & \sum\nolimits_{j\in\mathcal{J}_k}R_{j\to j}^k\\
			& \operatorname{s.t.} &{\rm (\ref{C_power allocation}),(\ref{C_Rmin}). }
		\end{eqnarray}
	\end{subequations}
	
	We divide the power allocation coefficient $p_{kj}$ into two parts, $\bar{p}_{kj}$ and $\triangle p_{kj}$, where $p_{kj}=\bar{p}_{kj}+\triangle p_{kj}$, with $\bar{p}_{kj}$ denoting the minimum power allocation coefficient for user $U_{kj}$ to satisfy the QoS constraint (\ref{C_Rmin}) and $\triangle p_{kj}$ denoting the power increment allocated to user $U_{kj}$. 
	Then, based on SIC decoding, the optimal power allocation coefficients can be obtained by the following lemma and proposition.

	\begin{lemma}
		\label{lemma-2}
		\emph{Problem (\ref{P_p-1}) is feasible if the following inequality holds:
			\begin{align}
				\label{p_feasible}
				\!\!\sum\nolimits_{j\in\mathcal{J}_k}\bar{p}_{kj}=\sum\nolimits_{j\in\mathcal{J}_k} \left(\prod\nolimits_{i=1}^{j-1}(r_{ki}^{\min}+1)\right)\frac{r_{kj}^{\min}}{\gamma_{kj}}\leq 1, 
			\end{align}
			where
			\begin{subequations}
			\begin{align}
				&r_{kj}^{\min}=2^{R_{kj}^{\rm min}}-1, 
				~\prod\nolimits_{i=1}^{0}(r_{kj}^{\min}+1)=1, \\ 
				& {and}~\gamma_{kj}=\frac{|\bar{\mathbf{h}}_{kj}\mathbf{f}_{k}|^2} 
				{|\bar{\mathbf{h}}_{kj}\mathbf{f}_{\bar{k}}|^2+\sigma_s^2\lVert\mathbf{g}_{kj}^{\mathrm H}\boldsymbol{\Theta}_{k}\lVert^2+\sigma_u^2}.
			\end{align}
	\end{subequations} } 
	\end{lemma}
	\begin{IEEEproof}
		See Appendix \ref{proof_of_lemma_2}.
	\end{IEEEproof}
	
	\begin{proposition}
		\label{proposition-1}
		\emph{If Lemma \ref{lemma-2} is satisfied, then the optimal power allocation coefficients are given by
			\begin{align}
				\label{optimal_p}
			\!\!\!\!\!\!\!\!\!	p_{kj}^{\star} \!= \! \left\{\begin{array}{l} \!\!\!
					\underbrace{\frac{r_{kj}^{\min}}{\gamma_{kj}}  + r_{kj}^{\min} \sum\nolimits_{i=j+1}^{J_k}\bar{p}_{ki}^{\star}}_{\bar{p}_{kj}^{\star}}  + \underbrace{r_{kj}^{\min}\sum\nolimits_{i=j+1}^{J_k}\triangle {p}_{ki}^{\star}}_{\triangle p_{kj}^{\star}},  \\
					~~~~~~~~~~~~~~~~~~~~~~~~~~~~~~~j  =1, 2,  \cdots, J_k  -  1, \\
					 \!\!\! \underbrace{\frac{r_{kj}^{\min}}{\gamma_{kj}}}_{\bar{p}_{kj}^{\star}}  +  \underbrace{\frac{1-\sum\nolimits_{i\in\mathcal{J}_k}\bar{p}_{ki}^{\star}}{\prod_{i=1}^{J_k-1}(1+r_{ki}^{\min})}}_{\triangle p_{kj}^{\star}},  ~j = J_k, 
				\end{array}\right.  \!\!\!\!\!\!\!\!\!
			\end{align}
			and the optimal objective value of (\ref{P_p-1}) is given by
			\begin{align}
				\nonumber
				&\sum\nolimits_{j\in\mathcal{J}_k} \log_2\left(1+\frac{ \bar{p}_{kj}^{\star}+r_{kj}^{\min} \sum\nolimits_{i=j+1}^{J_k}\triangle p_{ki}^{\star} }{\sum\nolimits_{i=j+1}^{J_k}{p}_{ki}^{\star} + \frac{1}{\gamma_{kj}}}\right) \\
				\label{optimal_R_j}
				&+\log_2\left(1+\frac{(1-\sum\nolimits_{j\in\mathcal{J}_k}\bar{p}_{kj}^{\star})\gamma_{k J_k}}{\prod_{i=1}^{J_k}(1+r_{ki}^{\min})}\right).
			\end{align}
		}
	\end{proposition}
	\begin{IEEEproof}
		See Appendix \ref{proof_of_proposition_1}.
	\end{IEEEproof}
	
	Proposition \ref{proposition-1} shows that the optimal power allocation coefficient, $p_{kj}^{\star}$, can be divided into two parts, $\bar{p}_{kj}^{\star}$ and $\triangle p_{kj}^{\star}$, where $p_{kj}^{\star}=\bar{p}_{kj}^{\star}+\triangle p_{kj}^{\star}$.
	Specifically, $\bar{p}_{kj}^{\star}$ maintains the QoS constraint of user $U_{kj}$, $\triangle p_{kJ_k}^{\star}$ maximizes the data rate of user $U_{kJ_k}$, while for other users, $\triangle p_{kj}^{\star}$ compensates for the SINR loss caused by the SIC interference.
	This is due to the fact that:
	1) the users are ordered according to their equivalent channel gains;
	2) at the optimum, improving the rate of one user comes at the cost of decreasing the rate of other users.
	
	\subsection{Transmit Beamforming Optimization} \label{Subsection_f}
	With given $\{p_{kj},\boldsymbol{\Theta}_k,\mathbf{w}\}$, we aim to solve the transmit beamforming vector $\mathbf{f}_k$.
	Problem (\ref{P1}) is still difficult to solve directly due to the non-concave objective function (\ref{P1_function}) and the non-convex constraint (\ref{C_Rmin}).
	To this end, we introduce auxiliary variables $Q_{kj}$ and $C_{kj}$, satisfying $Q_{kj}=R_{j\to j}^k$ and $C_{kj}=|\bar{\mathbf{h}}_{kj}\mathbf{f}_{k}|^2P_{kj}+B_{kj}$.
	The objective function (\ref{P0_function}) is then transformed into
	\begin{align}
		\label{R_C_P_f_1}
		\sum\nolimits_{k\in\mathcal{K}}\sum\nolimits_{j\in\mathcal{J}_k}R_{j\to j}^k=\sum\nolimits_{k\in\mathcal{K}}\sum\nolimits_{j\in\mathcal{J}_k} Q_{kj}.
	\end{align}
	In addition, we obtain the following new constraints:
	\begin{subequations}
		\begin{align}
			\label{R_C_P_f_2-1}
			&C_{kj}\geq |\mathbf{h}_{kj}\mathbf{f}_{k}|^2P_{kj}+B_{kj}, \\
			\label{R_C_P_f_2-2}
			&Q_{kj} \leq \log_2\big(1+p_{kj}A_{kj}^{-1}C_{kj}^{-1}\big), 
			~Q_{k j} \geq R_{kj}^{\rm min}.
		\end{align}
	\end{subequations}
    The SCA technique is employed to handle the non-convex constraint $Q_{kj} \leq \log_2\big(1+p_{kj}A_{kj}^{-1}C_{kj}^{-1}\big)$. 
    Specifically, a lower bound of its RHS in the $\ell$-th iteration is expressed as 
    \begin{align}
    \nonumber
    R_{kj}^{\rm lb}=&\log_2\Big(1+\frac{p_{kj}}{A_{kj}^{(\ell)}C_{kj}^{(\ell)}}\Big)-\frac{p_{kj}(\log_2 e)(A_{kj}-A_{kj}^{(\ell)})}{p_{kj}A_{kj}^{(\ell)}+(A_{kj}^{(\ell)})^2C_{kj}^{(\ell)}} \\
    \label{R_C_P_f_3}
    &-\frac{p_{kj}(\log_2 e)(C_{kj}-C_{kj}^{(\ell)})}{p_{kj}C_{kj}^{(\ell)}+(C_{kj}^{(\ell)})^2A_{kj}^{(\ell)}}.
    \end{align}
	Substituting (\ref{R_C_P_f_1})-(\ref{R_C_P_f_3}) into (\ref{P1}), the transmit beamforming optimization problem is written as
	 \begin{subequations}
		\label{P_f-0}
		\begin{align}
			\!\!\!\underset{\mathbf{f}_k,\Delta_1}{\max}  ~& \sum\nolimits_k\sum\nolimits_{j\in\mathcal{J}_k}Q_{kj}\\
			\operatorname{s.t.} ~& A_{kj}^{-1}\leq|\bar{\mathbf{h}}_{kj}\mathbf{f}_{k}|^2,	~\forall k\in\mathcal{K},\forall j\in\mathcal{J}_k, \\
			\nonumber
			& B_{kj}\geq |\bar{\mathbf{h}}_{kj}\mathbf{f}_{\bar{k}}|^2 +\sigma_s^2\lVert\mathbf{g}_{kj}^{\mathrm H}\boldsymbol{\Theta}_{k}\lVert^2 \\
			&~~~~~~~~+\sigma_u^2, 
			~\forall k\in\mathcal{K},\forall j\in\mathcal{J}_k,\\
			&C_{kj}\geq |\mathbf{h}_{kj}\mathbf{f}_{k}|^2P_{kj}+B_{kj}, 	~\forall k\in\mathcal{K},\forall j\in\mathcal{J}_k,\\
             \nonumber
            &\Gamma_{kj}\geq  A_{kj}^{-1}B_{kj}^{-1},
            ~\Gamma_{kj} \leq \Gamma_{kl}^{\rm lb}, \\
            \label{C_f_Gamma-1}
            &\forall k\in\mathcal{K},\forall j\in\mathcal{J}_k,\forall l\in\mathcal{L}_k,\\
            \label{C_f_Gamma-2}
            & Q_{kj}\!\leq \!R_{kj}^{\rm lb}, Q_{kj}\!\geq\! R_{kj}^{\min}, ~\forall k\!\in\!\mathcal{K},\forall j\!\in\!\mathcal{J}_k,\forall l \!\in\!\mathcal{L}_k,\!\!\!\!\\
            \label{C_f_mathcalC}
            &
            \mathcal{C}_m \geq 1+\exp\left(-a\left(\zeta_m-q\right)\right), ~\forall m,\\
            &\bar{W}\geq  P_{\rm O},~\zeta_m \leq P_{m}^{\rm RF}, ~\forall m,~ {\rm (\ref{C_transmit beamforming})},
		\end{align}
	\end{subequations}
    where $\Delta_1=\{A_{kj},B_{kj},C_{kj},Q_{kj},\Gamma_{kj},\mathcal{C}_m,\zeta_m\}$.
	We define $\bar{\mathbf{H}}_{kj}\!=\!\bar{\mathbf{h}}_{kj}^{\rm H}\bar{\mathbf{h}}_{kj}$ and $\mathbf{F}_k\!=\!\mathbf{f}_k\mathbf{f}_k^{\rm H}$, satisfying $\mathbf{F}_k\succeq \mathbf{0}$ and ${\rm rank}(\mathbf{F}_k)\!=\!1$.
	Then, problem (\ref{P_f-0}) is transformed	into
   	\begin{subequations}
   	\label{P_f-1}
   	\begin{align}
   		\!\!\!\!\underset{\mathbf{F}_k,\Delta_1}{\max}  ~&\sum\nolimits_{k\in\mathcal{K}}\sum\nolimits_{j\in\mathcal{J}_k}Q_{kj}\\
        \label{C_f_AB-1}
   		\operatorname{s.t.}~ 
   		&{A_{kj}^{-1}} \leq {\rm Tr}\left(\bar{\mathbf{H}}_{kj}\mathbf{F}_k\right),  ~\forall k\in\mathcal{K},\forall j\in\mathcal{J}_k,\\
   		\nonumber
   		& B_{kj} \geq {\rm Tr}\left(\bar{\mathbf{H}}_{kj}\mathbf{F}_{\bar{k}}\right)
   		+\sigma_s^2\lVert\mathbf{g}_{kj}^{\mathrm H}\boldsymbol{\Theta}_{k}\lVert^2 \\
   		\label{C_f_AB-2}
   		&~~~~~~~~+\sigma_u^2, ~\forall k\in\mathcal{K},\forall j\in\mathcal{J}_k,\\
   		\label{C_f_C}
   		&
   		C_{kj}\geq {\rm Tr}\left(\bar{\mathbf{H}}_{kj}\mathbf{F}_k\right)P_{kj} \!+\!B_{kj},~\forall k \! \in \!\mathcal{K},\forall j\in\mathcal{J}_k,\!\!\!\!\\
         \nonumber
   		& 	
   		\bar{W}\geq  \sum\nolimits_{k\in\mathcal{K}} \Big( \sum\nolimits_{k'\in\mathcal{K}} \operatorname{Tr} \left( \boldsymbol{\Theta}_{k} \mathbf{H} \mathbf{F}_{k'} \mathbf{H}^{\rm H} \boldsymbol{\Theta}_{k}^{\rm H} \right) \\
   		\label{C_f_P_OUT}
   		&~~~~~~~~~~~~~~~~~~~~~~~~~~~~~~~~~+ \sigma_s^2 \lVert \boldsymbol{\Theta}_{k} \lVert_F^2 \Big),\\
   		\nonumber
   		&\zeta_m  \leq\sum\nolimits_{k\in\mathcal{K}} \operatorname{Tr} \left( \bar{\mathbf{T}}_m\mathbf{H}\mathbf{F}_k\mathbf{H}^{\rm H}\bar{\mathbf{T}}_m^{\rm H}\right)\left(1-\alpha_{m}\right)\\
   		\label{C_f_P_RF}
   		&~~~~~~~~~~~~~~~~+\sigma_s^2\left(1-\alpha_{m}\right), ~\forall m,\\
   		\label{C_f_rank}
   		& 	{\rm rank}(\mathbf{F}_k)=1, ~\forall k,\\
   		 \label{C_f_Tr_F}
   		&\sum\nolimits_{k\in\mathcal{K}}\!\! {\rm Tr}\left(\mathbf{F}_k\right) \!\leq \!P_{\rm BS}^{\max}, \mathbf{F}_k\!\succeq \!0,  \forall k,{\rm (\ref{C_f_Gamma-1}){\text -}(\ref{C_f_mathcalC})},\!\!\!
   	\end{align}
   \end{subequations}
   where $\bar{\mathbf{T}}_m={\rm diag}([\underbrace{0,\cdots,}_{1 \ \text{to} \ m-1} 1\underbrace{,\cdots, 0}_{m+1 \ \text{to} \ M}])$.
    
    	\begin{algorithm}[tbp]
    	\setstretch{1}
    	\caption{The SROCR-Based Algorithm for Solving (\ref{P_f-1})}
    	\label{Algorithm_active}
    	\begin{algorithmic}[1] 
    		\STATE  Initialize feasible points $\{\mathbf{F}_{k}^{(0)}, w_{k}^{(0)}\}$ and the step size $\delta_1^{(0)}$. 
    		Set the iteration index $\ell_1=0$.
    		\REPEAT
    		\IF {problem (\ref{P_f-2}) is solvable} 
    		\STATE Update $\mathbf{F}_{k}^{(\ell_1+1)}$ by solving problem (\ref{P_f-2}); \STATE Update $\delta_1^{(\ell_1+1)}=\delta_1^{(\ell_1)}$;
    		\ELSE
    		\STATE  Update $\delta_1^{(\ell_1+1)}=\frac{\delta_1^{(\ell_1)}}{2}$;
    		\ENDIF 
    		\STATE Update $w_{k}^{(\ell_1+1)}=\min\Big(1,\frac{\lambda_{\rm max}(\mathbf{F}_{k}^{(\ell_1+1)})}{{\rm{Tr}}(\mathbf{F}_{k}^{(\ell_1+1)})}+\delta_1^{(\ell_1+1)}\Big)$; Update $\ell_1=\ell_1+1$;
    		\UNTIL the stopping criterion is met.
    	\end{algorithmic}  
    \end{algorithm}
    
   Next, we adopt the SROCR method to handle the rank-one constraint (\ref{C_f_rank}).
   Different from the conventional semidefinite relaxation (SDR) method that drops the rank-one constraint directly\cite{SPM-SDR}, the basic idea of SROCR is to relax the rank-one constraint gradually\cite{SROCR-EUS}.
	Specifically, we define $w_{k}^{(\ell-1)}\in[0,1]$ as the trace ratio parameter of $\mathbf{F}_{k}$ in the $(\ell\!-\!1)$-th iteration. 
	 Then, (\ref{C_f_rank}) in the $\ell$-th iteration is replaced by:
	\begin{eqnarray}
		\label{SROCR-linear-f}
		\big(\mathbf{f}_{k}^{{\rm e},(\ell-1)}\big)^{\rm H}\mathbf{F}_{k}^{(\ell)}\mathbf{f}_{k}^{{\rm e},(\ell-1)}\geq w_{k}^{(\ell-1)}{\rm Tr}(\mathbf{F}_{k}^{(\ell)}), ~\forall k, 
	\end{eqnarray}
	where $\mathbf{f}_{k}^{{\rm e},(\ell-1)}$ is the eigenvector corresponding to the largest eigenvalue of $\mathbf{F}_{k}^{(\ell-1)}$, and $\mathbf{F}_{k}^{(\ell-1)}$ is the solution obatined in the $(\ell-1)$-th iteration with given $w_{k}^{(\ell-1)}$. 
	Then, (\ref{P_f-1}) is reformulated as
	\begin{subequations}
		\label{P_f-2}
		\begin{eqnarray}
			&\underset{\mathbf{F}_k,\Delta_1}{\max}  &\sum\nolimits_{k\in\mathcal{K}}\sum\nolimits_{j\in\mathcal{J}_k}Q_{kj}\\
			&\operatorname{s.t.} & {\rm  (\ref{C_f_AB-1}){\text -}(\ref{C_f_P_RF}), (\ref{C_f_Tr_F}), (\ref{SROCR-linear-f})}.
		\end{eqnarray}
	\end{subequations}
	Problem (\ref{P_f-2}) is a convex semi-definite programming (SDP) problem, which can be solved efficiently via CVX\cite{CVX}. 
	By increasing $w_{k}^{(\ell-1)}$ from $0$ to $1$ over iterations, we approach a rank-one solution gradually.
	The algorithm for solving (\ref{P_f-1}) is given in Algorithm \ref{Algorithm_active}.
	After solving (\ref{P_f-1}), the solution of $\mathbf{f}_{k}$ is obtained by Cholesky decomposition of $\mathbf{F}_{k}$, e.g., $\mathbf{F}_{k}=\mathbf{f}_{k}\mathbf{f}_{k}^{\rm H}$.

	\subsection{MF-RIS Coefficient Deign}\label{MF-RIS Coefficients Deign}
	For any given $\{p_{kj},\mathbf{f}_{k},\mathbf{w}\}$, we define $\mathbf{U}_k\!=\!\mathbf{u}_k\mathbf{u}_k^{\rm H}$ and  $\mathbf{u}_k\!=\![\alpha_1\sqrt{\beta_1^k}e^{-j\theta_1^k};\!\cdots\!;\alpha_M\sqrt{\beta_M^k}e^{-j\theta_M^k};1\big]$, satisfying $\mathbf{U}_k\!\succeq\! \mathbf{0}$, ${\rm rank}(\mathbf{U}_k)\!=\!1$,  $[\mathbf{U}_{k}]_{m, m}\!=\!\alpha_m^2\beta_{m}^{k}$, and $[\mathbf{U}_{k}]_{M+1, M+1}\!=\!1$.
	Then, we have
	\begin{subequations}
		\begin{align}
			\label{C_A_H_passive-1}
			&|\bar{\mathbf{h}}_{kj}\mathbf{f}_{kj}|^2	=\operatorname{Tr}\big(\widetilde{\mathbf{H}}_{kj}\mathbf{F}_k\widetilde{\mathbf{H}}_{kj}^{\rm H}\mathbf{U}_{k}\big), \\
			\label{C_A_H_passive-2}
			&\lVert\mathbf{g}_{kj}^{\mathrm H}\boldsymbol{\Theta}_{k}\lVert^2=\operatorname{Tr}(\widetilde{\mathbf{G}}_{kj}\mathbf{U}_k),
			~P_{\rm O}=\sum\nolimits_{k\in\mathcal{K}}\operatorname{Tr}(\widetilde{\mathbf{H}}\mathbf{U}_{k}),
		\end{align}
	\end{subequations}
   where 	
		\begin{align}
			\nonumber
			&\widetilde{\mathbf{H}}_{kj}=\big[{\rm diag}(\mathbf{g}_{kj}^{\rm H})\mathbf{H}; \mathbf{h}_{kj}^{\rm H}], \\
			\nonumber
			&\widetilde{\mathbf{G}}_{kj}=[{\rm diag}(\mathbf{g}_{kj}^{\rm H}); \mathbf{0}_{1\times M}][{\rm diag}(\mathbf{g}_{kj}^{\rm H}); \mathbf{0}_{1\times M}]^{\rm H},\\
			\nonumber
			&\widetilde{\mathbf{H}}=\!\!\sum\nolimits_{k'\in\mathcal{K}} [\mathbf{H}\mathbf{f}_{k'}; 0][\mathbf{H}\mathbf{f}_{k'}; 0]^{\rm H} \!+\! \sigma_s^2 [\mathbf{I}_M;\mathbf{0}_{1\times M}] [\mathbf{I}_M;\mathbf{0}_{1\times M}]^{\rm H}.
		\end{align}
	Constraints (\ref{C_f_AB-1})-(\ref{C_f_P_OUT}) are then, respectively, rewritten as
	\begin{subequations}
		\label{C_theta_ABCP}
		\begin{align}
		&{A_{kj}^{-1}} \leq {\rm Tr}(\widetilde{\mathbf{H}}_{kj}\mathbf{F}_k\widetilde{\mathbf{H}}_{kj}^{\rm H}\mathbf{U}_k), \\
		&B_{kj} \geq{\rm Tr}\big((\widetilde{\mathbf{H}}_{kj}\mathbf{F}_{\bar{k}}\widetilde{\mathbf{H}}_{kj}^{\rm H}+\sigma_s^2\widetilde{\mathbf{G}}_{kj})\mathbf{U}_k\big)+\sigma_u^2, \\
		&C_{kj}\geq {\rm Tr}\big(\widetilde{\mathbf{H}}_{kj}\mathbf{F}_k\widetilde{\mathbf{H}}_{kj}^{\rm H}\mathbf{U}_k\big)P_{kj}+B_{kj},\\
		&\bar{\mathcal{W}}\geq \sum\nolimits_{k\in\mathcal{K}}\operatorname{Tr}( \widetilde{\mathbf{H}}\mathbf{U}_k).
		\end{align}
	\end{subequations}
   Accordingly, the MF-RIS coefficient design problem is formulated as
		\begin{subequations}
		\label{P_passive-1}
		\begin{eqnarray}
			\label{P_passive-1-objective}
			&\!\!\!\!\!\!\!\!\!\!\!\!\underset{\mathbf{U}_k,\Delta_1}{\max}  &\sum\nolimits_{k\in\mathcal{K}}\sum\nolimits_{j\in\mathcal{J}_k}Q_{kj}\\
			&\!\!\!\!\!\!\!\!\!\!\!\operatorname{s.t.} 
			\label{C_passive_rank_1}
			&\mathbf{U}_k\succeq 0, ~\left[\mathbf{U}_{k}\right]_{M+1, M+1}=1, ~\forall k,\\
			\label{C_passive_rank_2}
			&&\left[\mathbf{U}_{k}\right]_{m, m}=\alpha_m^2\beta_{m}^{k}, ~\forall m, k,\\
			\label{C_passive_rank_4}
			&&{\rm rank}(\mathbf{U}_k)=1, ~\forall k,\\
			\label{C_passive-alpha}
			&&\alpha_{m}\in\{0,1\}, ~\forall m,\\
			\label{C-passive-beta-1}
			&& \beta_{m}^k\in\left[0,\beta_{\max}\right], ~\sum\nolimits_{k\in\mathcal{K}} \beta_{m}^k\leq \beta_{\max}, ~\forall m, k, \\ 
			\label{C-passive-beta-2}
			&& {\rm (\ref{C_f_Gamma-1}){\text -}(\ref{C_f_mathcalC}), (\ref{C_f_P_RF}),  (\ref{C_theta_ABCP})}.
		\end{eqnarray}
	\end{subequations}

	The non-convexity of problem (\ref{P_passive-1}) arises from the non-convex constraint (\ref{C_passive_rank_2}), the rank-one constraint (\ref{C_passive_rank_4}), and the binary constraint (\ref{C_passive-alpha}).
	In Section \ref{Subsection_f}, we showed how to handle the rank-one constraint using SROCR.
	Similarly, by defining $v_{k}^{(\ell-1)}$, $\mathbf{u}_{k}^{{\rm e},(\ell-1)}$, and $\mathbf{U}_{k}^{(\ell)}$ to correspond to $w_{k}^{(\ell-1)}$, $\mathbf{f}_{k}^{{\rm e},(\ell-1)}$, and $\mathbf{F}_{k}^{(\ell)}$ in (\ref{SROCR-linear-f}), constraint (\ref{C_passive_rank_4}) is approximated by
	\begin{align}
		\label{SROCR-linear-theta}
		\big(\mathbf{u}_{k}^{{\rm e},(\ell-1)}\big)^{\rm H}\mathbf{U}_{k}^{(\ell)}\mathbf{u}_{k}^{{\rm e},(\ell-1)}
		\geq v_{k}^{(\ell-1)}{\rm Tr}(\mathbf{U}_{k}^{(\ell)}), ~\forall k.
	\end{align}
	
	The binary constraint (\ref{C_passive-alpha}) can be equivalently transformed into two continuous ones: $\alpha_m-\alpha_m^2\leq 0$ and $0\leq \alpha_m \leq 1$.
	However, $\alpha_m-\alpha_m^2\leq 0$ is still non-convex due to the non-convex term $-\alpha_m^2$.
	The SCA technique is employed to address it.
	Specifically, for a given point $\{\alpha_m^{(\ell)}\}$ in the $\ell$-th iteration, an upper bound is obtained as $\left(-\alpha_m^2\right)^{\rm ub}\!=\!-2\alpha_m^{(\ell)}\alpha_m\!+\!(\alpha_m^{(\ell)})^2$.
	
		\begin{algorithm}[tbp]
		\setstretch{1}
		\caption{The Penalty-Based Algorithm for Solving (\ref{P_passive-1})}
		\label{Algorithm_passive}
		\begin{algorithmic}[1] 
			\STATE  Initialize feasible points $\{\mathbf{U}_k^{(0)},v_k^{(0)}\}$, $\varepsilon>1$, and the step size $\delta_2^{(0)}$. 
			Set the iteration index $\ell_2=0$ and the maximum value of the penalty factor $\rho_{\max}$. 
			\REPEAT
			\IF {$\ell_2\leq \ell_2^{\max}$} 
			\IF {problem (\ref{P_passive-2}) is solvable} 
			\STATE Update $\mathbf{U}_k^{(\ell_2+1)}$ by solving problem (\ref{P_passive-2}); 
			\STATE Update $\delta_2^{(\ell_2+1)}=\delta_2^{(\ell_2)}$;
			\ELSE
			\STATE  Update $\delta_2^{(\ell_2+1)}=\frac{\delta_2^{(\ell_2)}}{2}$;
			\ENDIF
			\STATE Update ${v}_{k}^{(\ell_2+1)}=\min\Big(1,\frac{\lambda_{\rm max}(\mathbf{U}_{k}^{(\ell_2+1)})}{{\rm{Tr}}(\mathbf{U}_{k}^{(\ell_2+1)})}+\delta_2^{(\ell_2+1)}\Big)$;
			\STATE Update $\rho^{(\ell_2+1)}={\min}\{\varepsilon \rho^{(\ell_2)},\rho_{\max}\}$; 
			\STATE Update $\ell_2=\ell_2+1$;
			\ELSE
			\STATE Reinitialize with a new $\mathbf{U}_k^{(0)}$, set $\varepsilon>1$ and $\ell_2=0$.
			\ENDIF
			\UNTIL the stopping criterion is met.
		\end{algorithmic}  
	\end{algorithm}

	To address the highly-coupled constraint (\ref{C_passive_rank_2}), the auxiliary variable $\eta_m^k=\alpha_{m}^2\beta_{m}^k$ is introduced so that we can obtain the equivalent form of (\ref{C_passive_rank_2}) as
	\begin{align}
		\label{C_passive_rank_2-relax}
		\left[\mathbf{U}_{k}\right]_{m, m}=\eta_m^k, ~~\eta_m^k=\alpha_{m}^2\beta_{m}^k.
	\end{align}
   The non-convex constraint $\eta_m^k=\alpha_{m}^2\beta_{m}^k$ can be further transformed into the convex penalty term $G(\alpha_{m},\beta_{m}^k,\eta_m^k)$ $=\sum\nolimits_{k\in\mathcal{K}}\sum\nolimits_{m\in\mathcal{M}} (\frac{c_m^k}{2}\alpha_{m}^4+\frac{(\beta_{m}^k)^2}{2c_m^k}-\eta_m^k)$ by using the penalty-based method and convex upper bound (CUB) method, where the fixed point $\{c_m^k\}$ in the $\ell$-th iteration is updated by $(c_m^k)^{(\ell)}=\frac{(\beta_{m}^k)^{(\ell-1)}}{(\alpha_{m}^{(\ell-1)})^2}$; see Appendix \ref{proof_of_MF-RIS_coefficient} for the derivation details.
	Finally, problem (\ref{P_passive-1}) is recast as
	\begin{subequations}
		\label{P_passive-2}
		\begin{align}
			\label{P_passive-2-function}
			\underset{\mathbf{U}_k,\Delta_1,\eta_m^k}{\max}  ~&\sum\nolimits_{k\in\mathcal{K}}\sum\nolimits_{j\in\mathcal{J}_k}Q_{kj}-\rho G(\alpha_{m},\beta_{m}^k,\eta_m^k)  \\
			\operatorname{s.t.} ~&0\leq \alpha_m \leq 1, 
			~\alpha_m+\left(-\alpha_m^2\right)^{\rm ub}\leq 0, ~\forall m,\\
			&\left[\mathbf{U}_{k}\right]_{m, m}=\eta_m^k, ~\forall m, k,\\
			&{\rm  (\ref{C_passive_rank_1}), (\ref{C-passive-beta-1}), (\ref{C-passive-beta-2}),  (\ref{SROCR-linear-theta})},
		\end{align}
	\end{subequations}
	where the penalty factor $\rho>0$ penalizes the objective function (\ref{P_passive-2-function}) if $G(\alpha_{m},\beta_{m}^k,\eta_m^k)\neq 0$.
	It can be verified that, if $\rho\to\infty$, the solution obtained from (\ref{P_passive-2}) satisfies $G(\alpha_{m},\beta_{m}^k,\eta_m^k)=0$.
	Problem (\ref{P_passive-2}) is a convex SDP problem, which can be solved efficiently via CVX\cite{CVX}.
	The details of the proposed penalty-based algorithm are given in Algorithm \ref{Algorithm_passive}.

       \begin{figure*}[t]
    	\centering
    	\includegraphics[width=6.5in]{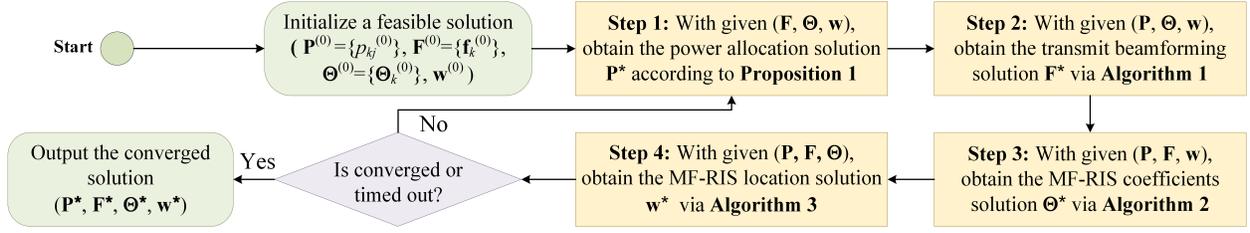}
    	\caption{A flowchart of the proposed AO algorithm.}
    	\label{flowchart}
    \end{figure*}      	
     
      \begin{algorithm}[tbp]
     	\setstretch{1}
     	\caption{The Local Region-Based Algorithm for Solving Problem (\ref{P_w-1})}
     	\label{Algorithm_location}
     	\begin{algorithmic}[1] 
     		\STATE Initialize feasible points $\{x^{(0)}, y^{(0)},z^{(0)},t^{(0)},t_{kj}^{(0)},\bar{v}^{(0)}\}$. 
     		Set the iteration index $\ell_3=0$.
     		\REPEAT  
     		\STATE Update $\{x^{(\ell_3\!+\!1)},y^{(\ell_3\!+\!1)},z^{(\ell_3\!+\!1)},t^{(\ell_3\!+\!1)},t_{kj}^{(\ell_3\!+\!1)},\bar{v}^{(\ell_3\!+\!1)}\}$ by solving problem (\ref{P_w-3});
     		\STATE Update $\ell_3=\ell_3+1$;
     		\UNTIL the stopping criterion is met.
     	\end{algorithmic}  
     \end{algorithm}
     
	\subsection{MF-RIS Location Optimization}
	Finally, we focus on the location optimization of the MF-RIS.
	Equations (\ref{H_bs}), (\ref{H_bs_los}), and (\ref{h_skj}) show that both the distance-dependent path loss, $L_{bs}$ and $L_{skj}$, and the LoS components, $\mathbf{H}^{\rm LoS}$ and $\mathbf{g}_{kj}^{\rm LoS}$, are relevant to the MF-RIS location, $\mathbf{w}$.
	In addition, (\ref{H_bs_los}) shows that these LoS components are non-linear w.r.t. $\mathbf{w}$, which are difficult to deal with directly.
	Here, we invoke the local region optimization method to tackle this issue\cite{SF-RIS-user fairness-3}.
	Denote $\mathbf{w}^{(i-1)}$ as the feasible location obtained in the $(i-1)$-th iteration, then the location variables should satisfy the following constraint:
	\begin{align}
		\label{C_local region}
		\big\|\mathbf{w}-\mathbf{w}^{(i-1)}\big\|\leq \epsilon,
	\end{align}
	where the constant $\epsilon$ is relatively small such that $\mathbf{w}^{(i-1)}$ can be used to approximately obtain $\mathbf{H}^{\rm LoS}$ and $\mathbf{g}_{kj}^{\rm LoS}$ in the $i$-th iteration.
	Assuming that $\hat{\mathbf{H}}^{(i-1)}$ and $\hat{\mathbf{g}}_{kj}^{(i-1)}$ are the small-scale fading obtained in the $(i-1)$-th iteration, constraints (\ref{C_f_AB-1})-(\ref{C_f_P_RF}) are, respectively, rewritten as 
	\begin{subequations}
		\label{C_appendix-D-ABC}
		\begin{align}
			\label{C_AB_w}
			&\!\!A_{kj}^{-1} \! \leq \! \mathbf{d}_{kj}^{\rm T}\mathbf{D}_{kj}\mathbf{d}_{kj}, 
			~B_{kj} \! \geq \! \mathbf{d}_{kj}^{\rm T}\bar{\mathbf{D}}_{kj}\mathbf{d}_{kj} 
		\!	+ \! d_{skj}^{-\kappa_{2}}\mathcal{W}_1 \! + \! \sigma_u^2, \! \! \\
			\label{C_C_w}
			&\!\!C_{kj}\! \geq \! \mathbf{d}_{kj}^{\rm T}\mathbf{D}_{kj}\mathbf{d}_{kj}P_{kj} \! + \! B_{kj},
             ~d_{bs}^{-\kappa_{0}} \! \leq \! \mathcal{W}_2, 
			~d_{bs}^{-\kappa_{0}} \! \geq \! \mathcal{W}_3,\!\!
		\end{align}
	\end{subequations}
	where $\mathbf{d}_{kj}=\big[1,  d_{bs}^{-\frac{\kappa_{0}}{2}}d_{skj}^{-\frac{\kappa_{2}}{2} }\big]^{\rm T}$. Here, $\mathbf{D}_{kj}$, $\bar{\mathbf{D}}_{kj}$, $\mathcal{W}_1$, $\mathcal{W}_2$, and $\mathcal{W}_3$ are terms unrelated to the optimization variable $\mathbf{w}$ in the $i$-th iteration, which are given by (\ref{W_independent}) in Appendix \ref{Parameters introduced in constraint}.
   Accordingly, given $\{p_{kj},\mathbf{f}_k,\boldsymbol{\Theta}_k\}$, problem (\ref{P1}) is reduced to
   	\begin{subequations}
   	\label{P_w-1}
   	\begin{align}
   		\underset{\mathbf{w},\Delta_1}{\max}  ~&\sum\nolimits_{k\in\mathcal{K}}\sum\nolimits_{j\in\mathcal{J}_k}Q_{kj}\\
   		\operatorname{s.t.}~&{\rm (\ref{MF-RIS region}), (\ref{C_f_Gamma-1}){\text -}(\ref{C_f_mathcalC}), (\ref{C_local region}), (\ref{C_AB_w}), (\ref{C_C_w})}.
   	\end{align}
   \end{subequations} 

	Constraints (\ref{C_AB_w}) and (\ref{C_C_w}) are still non-convex w.r.t. $\mathbf{w}$.
	To tackle this issue, we first introduce auxiliary variables to replace their complex terms, and then approximate the non-convex parts by using SCA.
    Specifically, by introducing a slack variable set $\Delta_2\!=\!\{t, t_{kj}, \bar{t}_{kj}, e_{kj},v,\bar{v}, r_{kj},\bar{r}_{kj}, s_{kj}\}$ and defining $\bar{\mathbf{d}}_{kj}=[1,\bar{t}_{kj}]^{\rm T}$, these constraints are linearly approximated by (\ref{Appendix-D-1}){\text -}(\ref{Appendix-D-2}), (\ref{Appendix-D-3}), and (\ref{P_w-C-SCA-appendix}) in Appendix  \ref{proof_of_lemma_3}. 
     
	Next, reformulating problem (\ref{P_w-1}) by replacing constraints (\ref{C_AB_w}) and (\ref{C_C_w}) with convex ones yields the following optimization problem:
	\begin{subequations}
		\label{P_w-3}
		\begin{eqnarray}
			&\!\!\!\!\!\!\!\!\!\!\!\!\!\!\underset{\mathbf{w},\Delta_1,\Delta_2}{\max}  &\!\!\!\!\sum\nolimits_{k\in\mathcal{K}}\sum\nolimits_{j\in\mathcal{J}_k}Q_{kj}\\
			&\!\!\!\!\!\!\!\!\!\!\!\!\!\!\operatorname{s.t.}&\!\!\!\! {\rm (\ref{MF-RIS region}), (\ref{C_f_Gamma-1}){\text -}(\ref{C_f_mathcalC}),(\ref{C_local region}), (\ref{Appendix-D-1}){\text -}(\ref{Appendix-D-2}), (\ref{Appendix-D-3}),  (\ref{P_w-C-SCA-appendix}).}
		\end{eqnarray}
	\end{subequations}
	Problem (\ref{P_w-3}) is convex, which can be solved efficiently via CVX\cite{CVX}.
	The details of the proposed local region-based algorithm are given in Algorithm \ref{Algorithm_location}.

	Based on the above solutions, a flowchart of the overall AO algorithm for solving problem (\ref{P0}) is given in Fig. \ref{flowchart}.
    Since the optimal power allocation is obtained in closed form in Proposition \ref{proposition-1}, the complexity of Step 1 is $\mathcal{O}(1)$.
	The complexity of the SDP problems in Steps 2 and 3 is  $\mathcal{O}_{\mathbf{F}}=\mathcal{O}\big(I_{\mathbf{F}}\max (2N, 3J+M)^4\sqrt{2N}\big)$ and $\mathcal{O}_{\boldsymbol{\Theta}}=\mathcal{O}\big(I_{\boldsymbol{\Theta}}\max (2(M+1), 3J)^4\sqrt{2(M+1)}\big)$, respectively, while the complexity of Step 4 using the interior-point method is $\mathcal{O}_{\mathbf{w}}=\mathcal{O}\big(I_{\mathbf{w}}(6+2M+11J)^{3.5}\big)$.
	Here, $I_{\mathbf{F}}$, $I_{\boldsymbol{\Theta}}$, and $I_{\mathbf{w}}$ represent the respective number of iterations\cite{SPM-SDR}.
	Since each sub-algorithm converges to a local optimum, the objective value of problem (\ref{P0}) is non-decreasing after each iteration.
	Moreover, the maximum transmit power constraint (\ref{C_transmit beamforming}) indicates that the objective value has an upper bound. 
	Hence, the AO algorithm is guaranteed to converge.

	\section{Numerical Results} \label{Numerical Results}	
\begin{figure}[t]
		\centering
		\includegraphics[width=3 in]{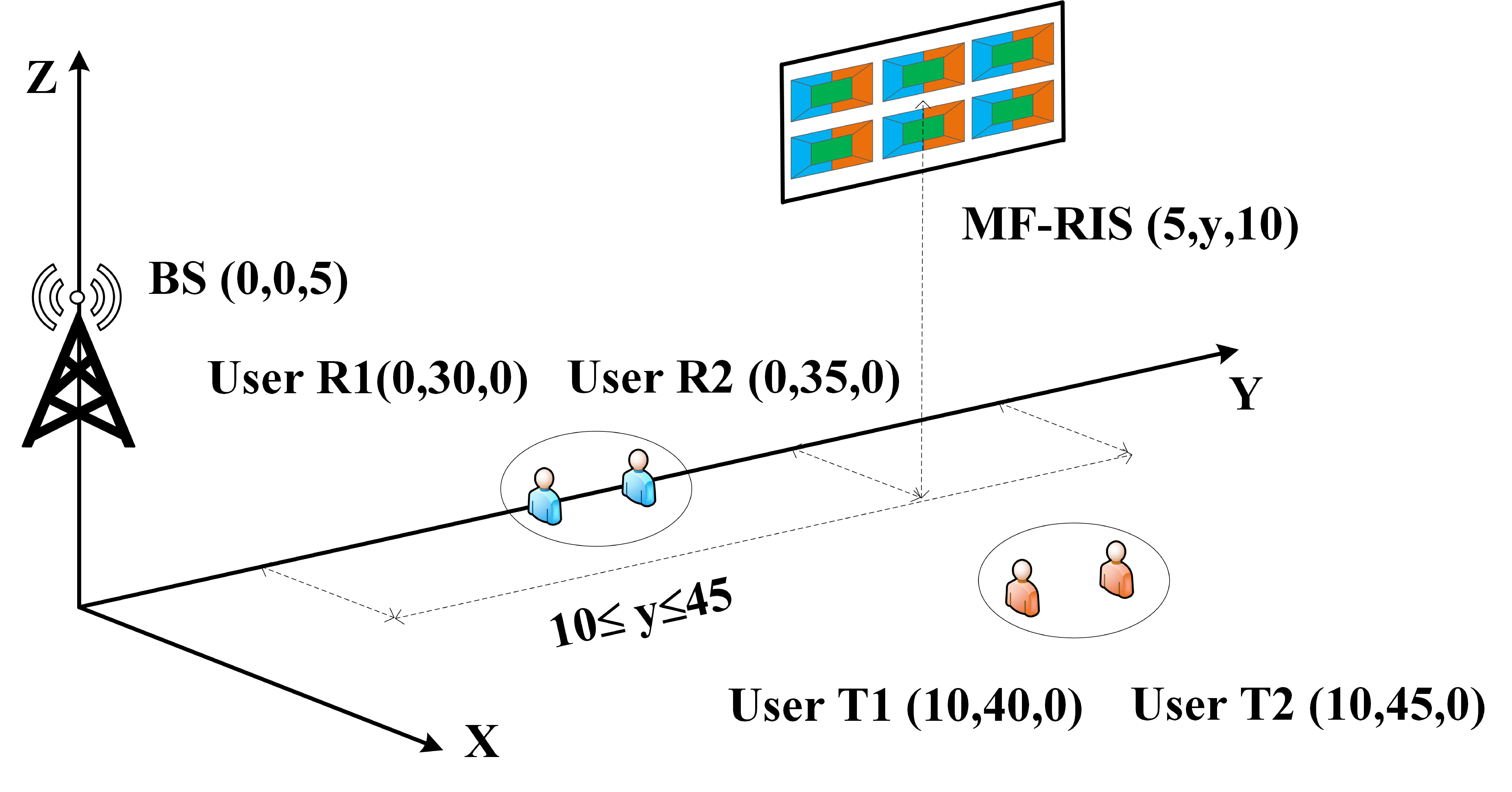}
		\caption{Simulation setup.}
		\label{simulation figure}
		\vspace{-3mm}
	\end{figure}

	In this section, numerical results are provided to validate the effectiveness of the proposed algorithm and the superiority of the considered MF-RIS assisted NOMA system over existing SF- and DF-RIS assisted systems.
	As shown in Fig. \ref{simulation figure}, we consider a scenario with $J_r=J_t=2$ users, where the BS is located at $\mathbf{w}_b = [0,0,5]^{\rm T}$ m and the MF-RIS deployable region is set as $\mathcal{P} = \{[5,y,10]^{\rm T}|10 \leq y  \leq 45\}$.
	All users are randomly distributed in their own circle with the radius of $2$ m.
	The corresponding centers are set as $[0,30,0]^{\rm T}$,  $[0,35,0]^{\rm T}$, $[10,40,0]^{\rm T}$, and $[10,45,0]^{\rm T}$ m, respectively.
	Unless otherwise specified, we set $P_{\rm BS}^{\max}=40$ dBm and $M=120$.
	Other parameters are summarized in Table \ref{Simulation parameters}.
	
	\begin{table}[t]
		\centering
		\renewcommand{\arraystretch}{0.95}
		\caption{Simulation parameters}
		\vspace{-2mm}
		\label{Simulation parameters}
		\scalebox{0.9}{
			\begin{tabular}{|c|c|}
				\hline  
				\bfseries Parameter & \bfseries Value\\
				\hline
				\tabincell{c}{ Communication \\ parameters } & \tabincell{c}{$h_0= -20$ dB\cite{sustainable-RIS-TC-WPT}, $\kappa_{0}=2.2$, $\kappa_{1}=2.8$,  \\ $\kappa_{2}=2.6$,  $\beta_{0}=\beta_{1}=\beta_{2}=3$ dB, \\  $\sigma_u^2=\sigma_s^2=-70$ dBm} \\
				\hline
				\tabincell{c}{ Power consumption \\ parameters }&  \tabincell{c}{$\xi=1.1$, $P_b=1.5$ mW, $P_{\rm DC}=0.3$ mW, \\ $P_{\rm C}=2.1$ $\mu$W\cite{sustainable-RIS-TCOM},  $Z=24$ mW, \\  $a=150$, $q=0.014$\cite{EH-parameter}} \\
				\hline
				Other parameters & \tabincell{c}{ $N=4$,  $\beta_{\max}=20$ dB\cite{active-RIS-TWC-EE}, \\ $\frac{d}{\lambda}=0.5$,  $\rho^{(0)}=10^{-3}$, $\epsilon=0.05$\cite{SF-RIS-user fairness-3} } \\
				\hline
		\end{tabular}}
		\vspace{-2mm}
	\end{table}

	\begin{table*}[t]
		\centering
		\renewcommand{\arraystretch}{1.1}
		\caption{Benchmark algorithms}
		\label{Comparison_algorithm}
		\vspace{-2mm}
		\scalebox{0.92}{
			\begin{tabular}{|c|c|c|c|c|c|}
				\hline  
				\bfseries Algorithm & \bfseries Power allocation & \bfseries Transmit beamforming & \bfseries MF-RIS coefficient & \bfseries MF-RIS location  \\
				\hline
				Exhaustive search-based algorithm& Exhaustive search & Algorithm 1 & Algorithm 2 & Exhaustive search \\
				\hline
				SDR-based algorithm & Proposition 1 & SDR & SDR & Algorithm 3  \\
				\hline
				MRT-based algorithm & Proposition 1 & MRT & Algorithm 2 & Algorithm 3  \\
				\hline
				Random-based algorithm & Proposition 1 & Algorithm 1 & Random coefficient & Algorithm 3  \\
				\hline
		\end{tabular}}
		\vspace{-2mm}
	\end{table*}

	To evaluate the performance of the proposed algorithm, we consider four benchmarks, as summarized in Table \ref{Comparison_algorithm}, i.e.:
	\begin{itemize}
	  \item {\bf Exhaustive search-based algorithm}: The power allocation factors and MF-RIS locations are optimized by the exhaustive search. This case can be regarded as providing a performance upper bound of our proposed algorithm.
	  
	  \item {\bf SDR-based algorithm:} The transmit beamforming and MF-RIS coefficients are designed by adopting the SDR method, which ignores the rank-one constraints (\ref{C_f_rank}) and (\ref{C_passive_rank_4})\cite{SPM-SDR}. The Gaussian randomization approach is applied when the obtained solution is not rank-one.
	  
	  \item {\bf MRT-based algorithm:} The transmit beamforming optimization problem is solved by invoking the maximum-ratio transmission (MRT) method\cite{SF-RIS-energy consumption-3}.
	  
	  \item {\bf Random-based algorithm:} The MF-RIS coefficients are randomly set within the feasible region $\mathcal{R}_{\rm MF}$.
	\end{itemize}
	
	    \begin{figure}[t]
		\centering
			\includegraphics[width=3.1in]{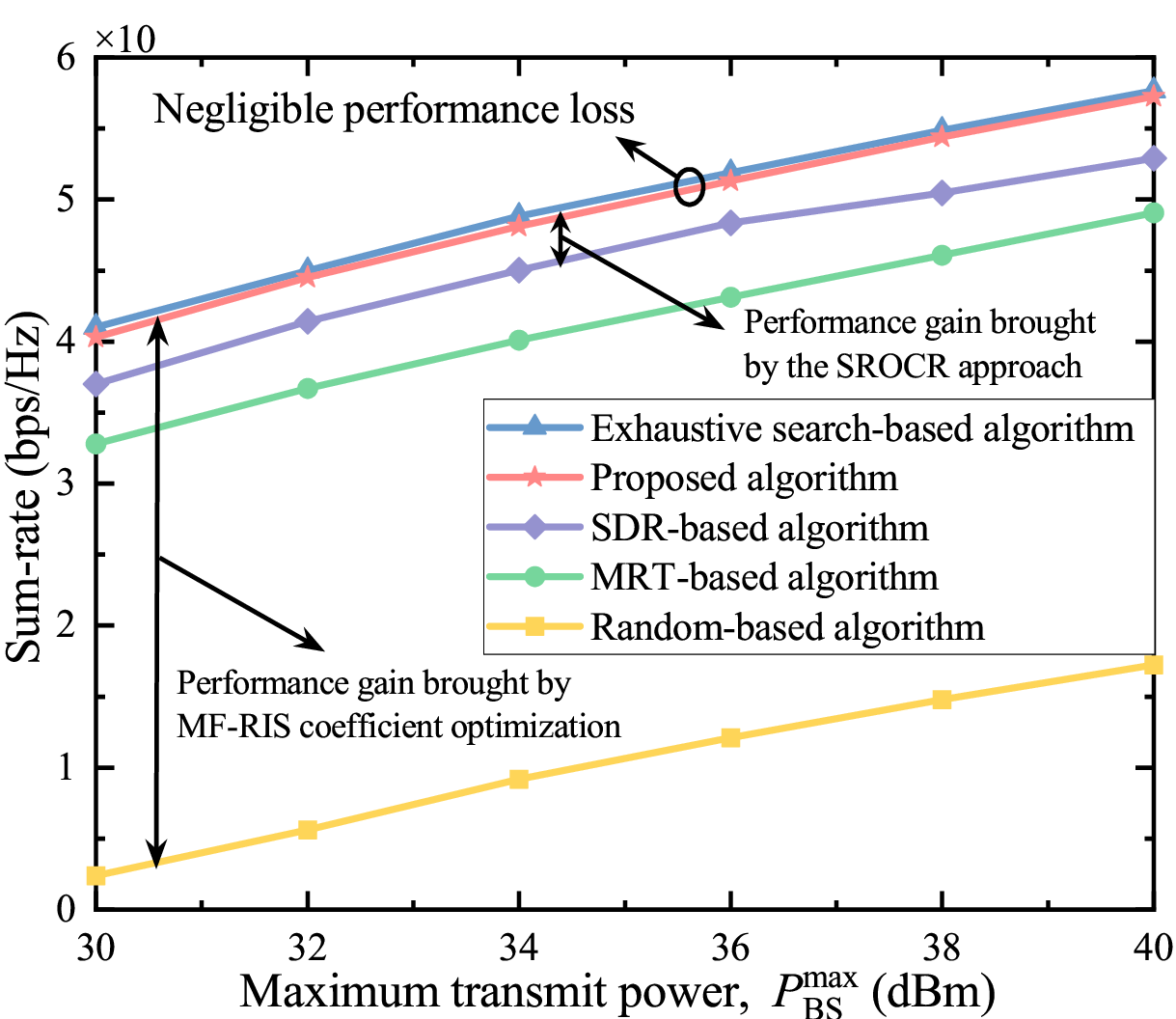}
		\vspace{-2mm}
		\caption{Sum-rate versus $P_{\rm BS}^{\max}$ under different algorithms.}
	\label{Algorithm}
	\vspace{-2mm}
	\end{figure}
	
	Fig. \ref{Algorithm} shows the sum-rate versus $P_{\rm BS}^{\max}$ under different algorithms. 
	The proposed algorithm achieves comparable performance to the exhaustive search-based algorithm with relatively low complexity. 
	Specifically, the complexity of the proposed power allocation and MF-RIS location optimization algorithms is $\mathcal{O}(1)$ and $\mathcal{O}\big(I_{\mathbf{w}}(6+2M+11J)^{3.5}\big)$, while the complexity of the exhaustive search with accuracy $\varsigma$ is $\mathcal{O}(\frac{1}{\varsigma^{J}})$ and $\mathcal{O}(\frac{1}{\varsigma^{3}})$, respectively.
	The random-based algorithm performs significantly worse than the proposed algorithm due to the non-optimized MF-RIS coefficients. 
	Besides, the proposed algorithm achieves a higher sum-rate gain than the SDR-based algorithm.
	This is because using the SDR method to solve the relaxed problem usually generates a high-rank solution, and the constructed solution is normally suboptimal or even infeasible for the original problem\cite{SROCR-EUS}. In contrast, the adopted SRCOR method can approach a locally optimal rank-one solution. 
	Additionally, it is observed that the proposed algorithm outperforms the MRT-based algorithm, which confirms the importance of joint optimization of the transmit beamforming and other variables.
	
     To demonstrate the benefits brought by the proposed MF-RIS, we consider the following schemes. Unless otherwise specified, the NOMA technique is adopted for all schemes.
	\begin{itemize}
		\item \textbf{MF-RIS with $\beta_{m}^k\in\mathcal{R}_{\rm MF}^{1}=\{\beta_m^k|\beta_m^k\in[0,\beta_{\rm max}],$ $\sum_{k}\beta_m^k \leq \beta_{\rm max},\forall m,k\}$}: The BS is assisted by the proposed MF-RIS, where the elements in S mode reflect, refract, and amplify the incident signal simultaneously.
		\item \textbf{MF-RIS with $\beta_{m}^k\in\mathcal{R}_{\rm MF}^{ 2}=\{\beta_m^k|\beta_m^k\in[0,\beta_{\max}],$ $\prod\nolimits_k\beta_m^k=0,\forall m,k\}$}: This scheme considers a special case of the MF-RIS in which the elements in S mode are divided into two groups.
        One group is used to serve users in the front half-space, while the other is used to serve users in the back half-space.
        This group-wise amplitude control reduces the overhead caused by exchanging configuration information between the BS and the MF-RIS, making it easier to implement in practical applications.
		
		\item \textbf{Active RIS\cite{active-RIS-analysis-1}}: This type of RIS can amplify and reflect signals simultaneously, or refract and amplify signals simultaneously, but cannot support energy harvesting.
		\item \textbf{STAR-RIS\cite{STAR-Liu}}: The communications from the BS to all users are assisted by a STAR-RIS, i.e., $\alpha_m=1,\beta_m^k\in[0,1],\sum\nolimits_k\beta_m^k\leq 1,\theta_m^k\in[0,2\pi),\forall m, k$.
		Compared to the proposed MF-RIS, the STAR-RIS does not support signal amplification and energy harvesting.
		\item \textbf{Self-sustainable RIS\cite{sustainable-RIS-TCOM}}: This type of RIS allows
		a portion of the elements to operate in signal reflection or refraction mode, while the rest work in H mode.
		\item \textbf{Conventional RIS\cite{SF-RIS-energy consumption-3}}: In this scheme, the RIS only supports reflection or refraction.
		\item \textbf{Without RIS}: This is a baseline with no RIS deployment. Only direct links are considered from the BS to the users.
	\end{itemize}
	
	To achieve full-space coverage, for the active RIS, self-sustainable RIS, and conventional RIS, one reflective RIS and one refractive RIS are deployed adjacent to each other at the same location as the MF-RIS, and each RIS has $M/2$ elements.
    In addition, for fairness, we define the total power budget as $P_{\rm total}^{\rm max}$, where $P_{\rm total}^{\rm max}=P_{\rm BS}^{\rm max}+P_{\rm RIS}^{\rm max}$ and  $P_{\rm BS}^{\rm max}=P_{\rm RIS}^{\rm max}$ hold for the active RIS-aided schemes, and $P_{\rm total}^{\rm max}=P_{\rm BS}^{\rm max}$ holds for other schemes.
    Here, $P_{\rm RIS}^{\rm max}$ denotes the maximum amplification power at the active RIS.
	
	   \begin{figure}[t]
		\centering
		\includegraphics[width=3in]{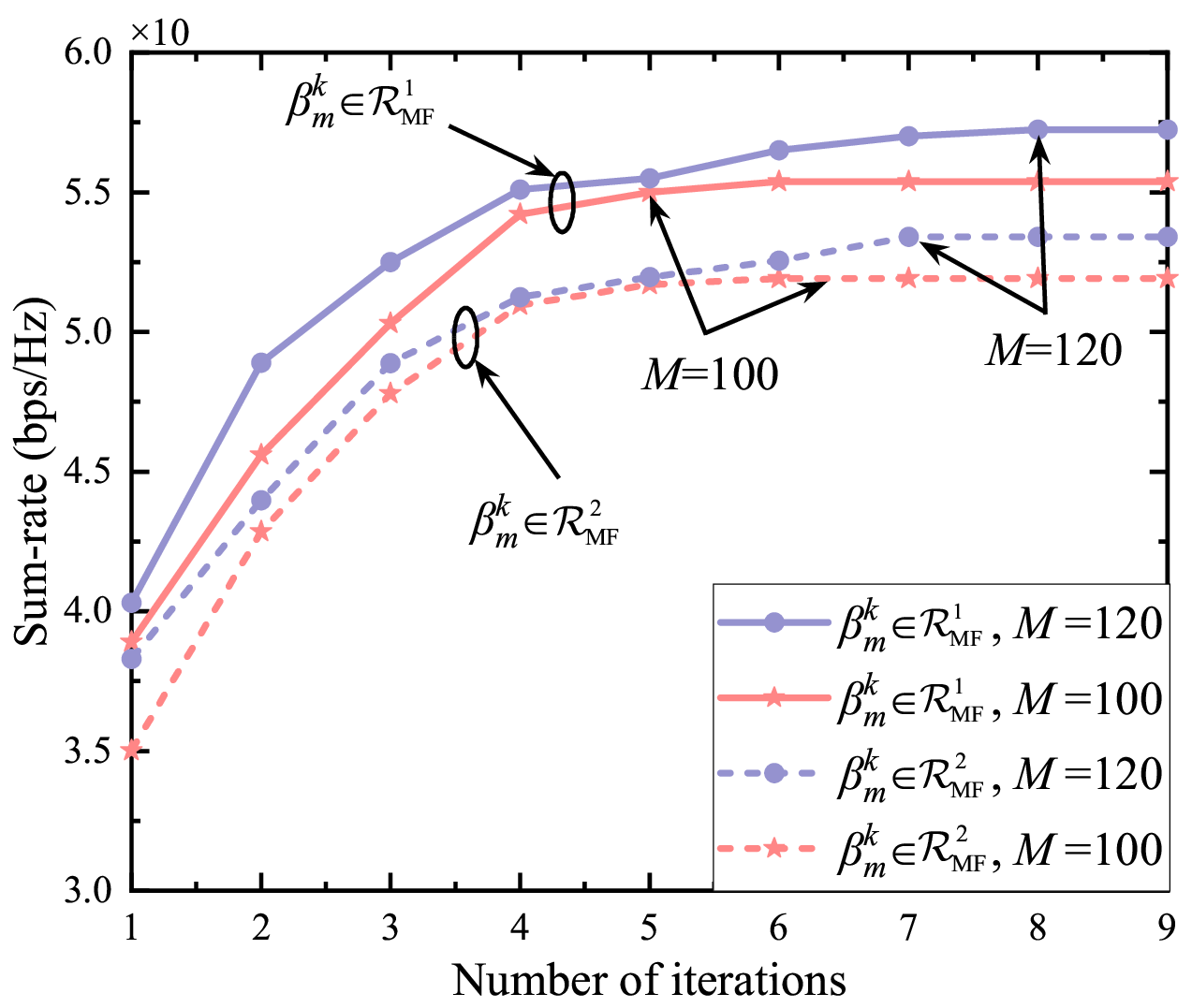}
		\vspace{-2mm}
		\caption{Convergence behavior of the proposed AO algorithm under different $M$ and different operating schemes.}
		\label{Convergence}
		\vspace{-2mm}
	\end{figure}

	The convergence behavior of the proposed algorithm with different numbers of MF-RIS elements and different operating schemes is illustrated in Fig. \ref{Convergence}.
	All curves gradually increase and exhibit the trend of convergence after a finite number of iterations.
	Specifically, the proposed algorithm with $M=100$ converges to a stable value after about $6$ iterations.
	However, for the cases with $M=120$, it requires around $8$ iterations for convergence.
	This is because both the number of optimization variables and the number of constraints increase with $M$, and thus increase the complexity of solving (\ref{P0}).

	\begin{figure}[t]
		\centering
	\includegraphics[width=3in]{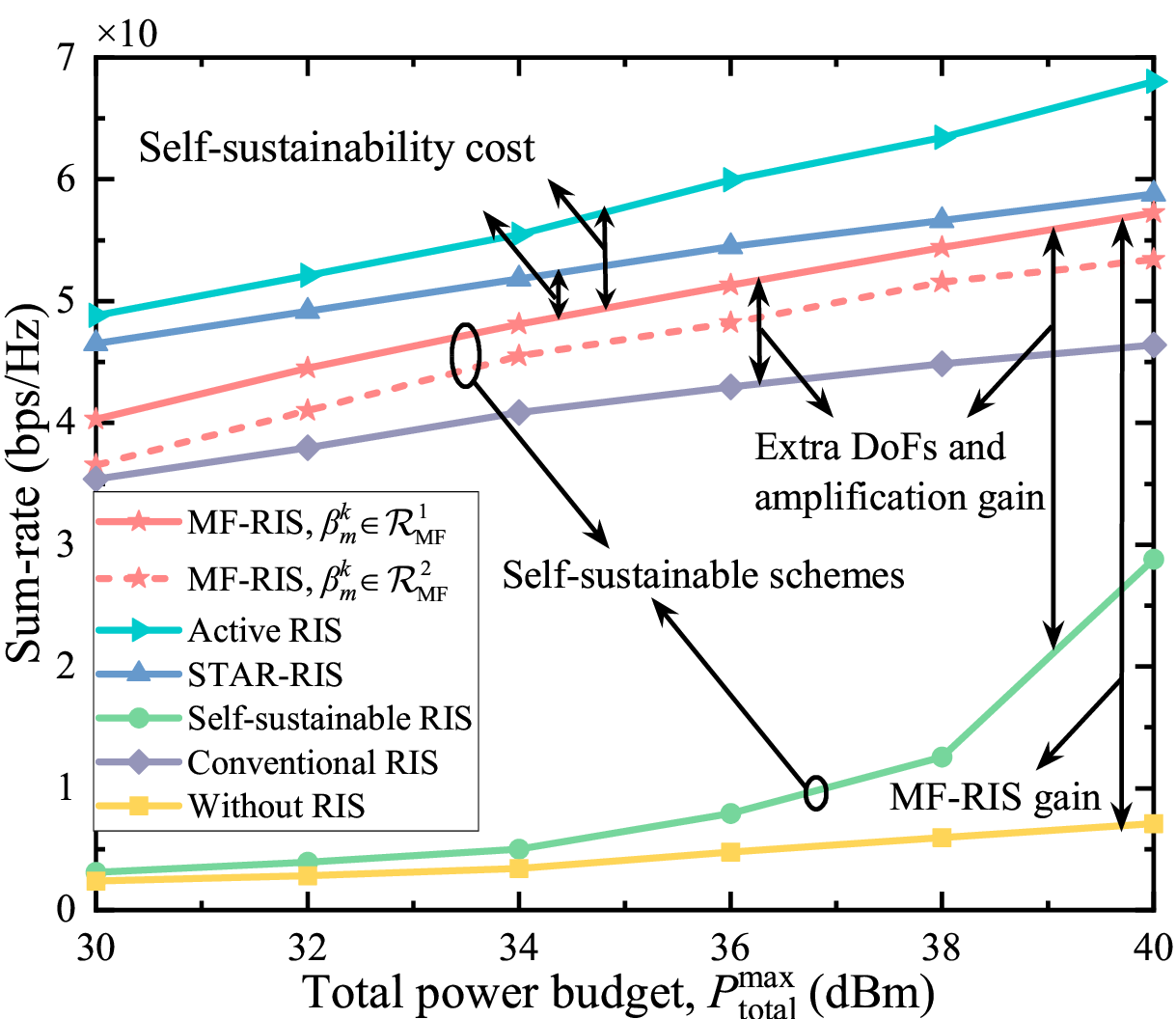}
		\vspace{-2mm}
	\caption{Sum-rate versus $P_{\rm total}^{\max}$ under different schemes.}
	\label{Pmax_STAR_MF}
	\vspace{-2mm}
	\end{figure}

   Fig. \ref{Pmax_STAR_MF} depicts the sum-rate versus the total power budget.
	For the two self-sustainable schemes, the proposed MF-RIS is far superior to the self-sustainable RIS.
	Specifically, when $P_{\rm total}^{\max}\!=\!40$ dBm, the MF-RIS scheme enjoys $98.8$\% higher sum-rate gain than the scheme with self-sustainable RIS.
	This result can be explained as follows.
	The MF-RIS adjusts its coefficients in an element-wise manner, while the self-sustainable RIS employs a fixed mode for each element. 
	The latter suffers from throughput loss due to its limited DoFs.
	Moreover, the signal amplification design effectively alleviates the impact of ``double attenuation" on the MF-RIS-aided links. 
	This means that the proposed MF-RIS makes better use of the harvested energy and thus increases the sum-rate of all users.
	Additionally, compared to the conventional RIS, the MF-RIS achieves $23.4$\% higher sum-rate gain due to its resource allocation flexibility and signal amplification function.
	
	We observe from Fig. \ref{Pmax_STAR_MF} that the throughput performance of active RIS and STAR-RIS is better than that of MF-RIS. 
	This observation can be explained as follows:
	1) the active RIS and the STAR-RIS use all elements to serve users, while the MF-RIS only uses part of the elements to relay signals;
	2) unlike the active RIS and the STAR-RIS, which assume an ideal lossless signal relay and power supply process, the MF-RIS takes into account the inevitable power loss and circuit consumption during energy harvesting and signal amplification.
   Although the self-sustainability of MF-RIS comes with the decreased performance, the sum-rate loss decreases with $P_{\rm total}^{\max}$ due to the fact that the elements in H mode can harvest more energy at high power.
   Besides, the active RIS is superior to the STAR-RIS and the gain increases with $P_{\rm total}^{\max}$, indicating that a larger RIS amplification power allows the signal amplification gain to be greater than the gain of full-space coverage.
	For the MF-RIS, the scheme of $\beta_{m}^k\in\mathcal{R}_{\rm MF}^{1}$ always performs better than the scheme of $\beta_{m}^k\in\mathcal{R}_{\rm MF}^{2}$.
	This is because the continuous set $\mathcal{R}_{\rm MF}^{1}$ allows for more flexibility in amplitude modeling, while the binary set $\mathcal{R}_{\rm MF}^{2}$ restricts the capacity of each element to enhance the desired signal, suppress inter-user interference, and ultimately reduces the achievable sum-rate.

			\begin{figure}[t]
		\centering
		\includegraphics[width=3in]{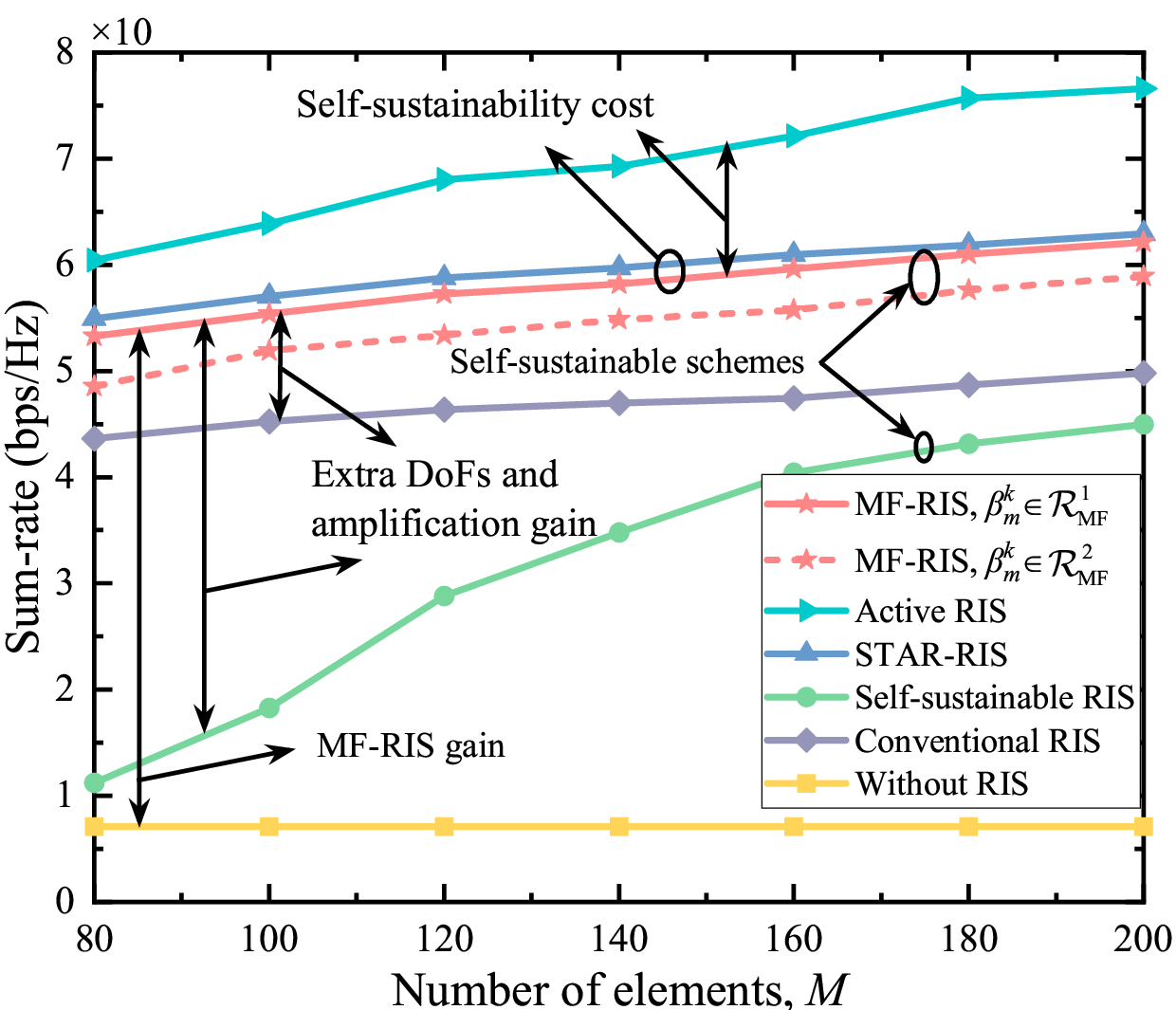}
		\vspace{-2mm}
		\caption{Sum-rate versus $M$ under different schemes.}
		\label{M_STAR_MF}
		\vspace{-2mm}
	\end{figure}
	
		\begin{figure}[t]
		\centering
		\includegraphics[width=3in]{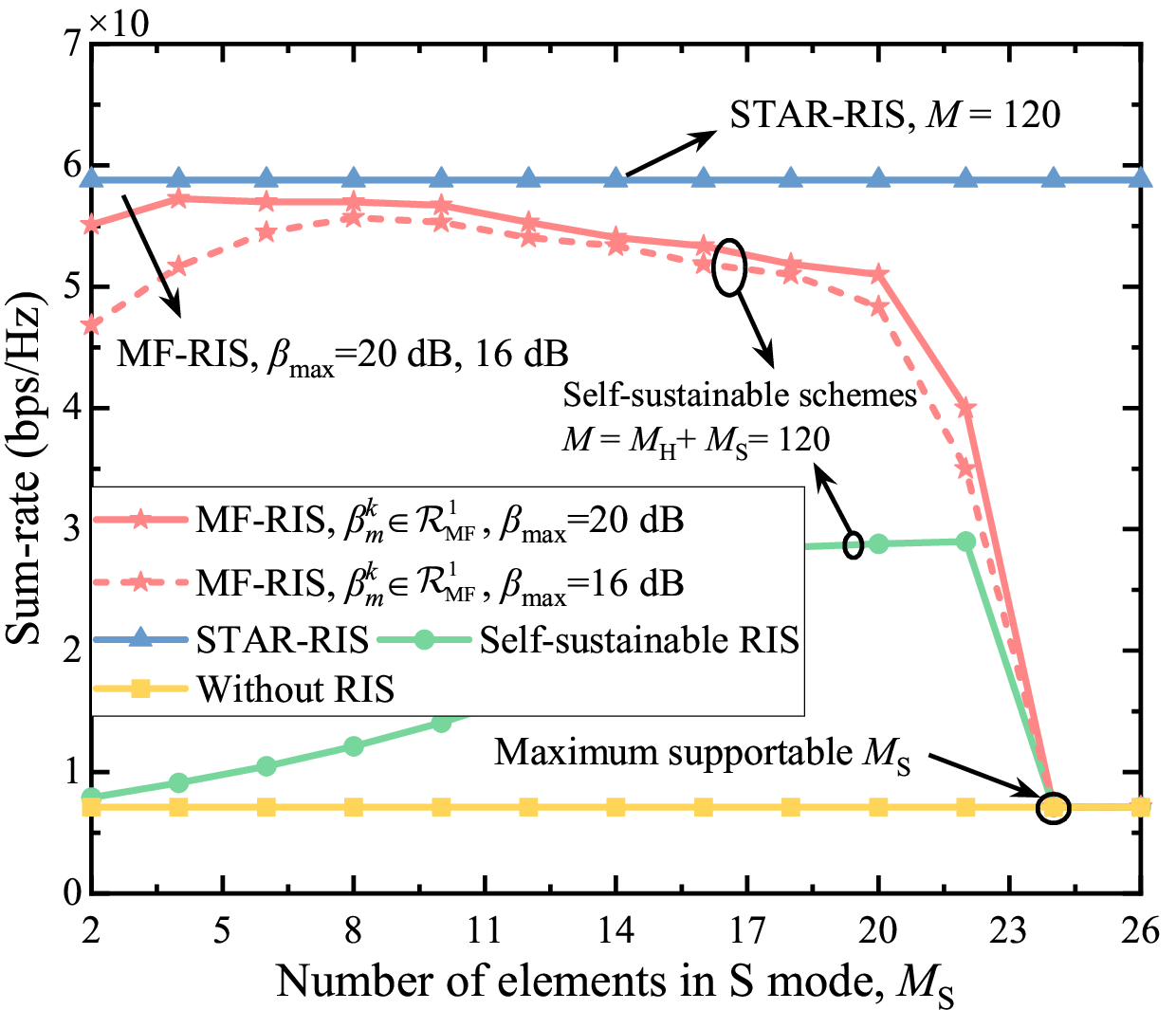}
		\vspace{-2mm}
		\caption{Sum-rate versus $M_{\rm S}$ under different schemes.}
		\label{Fixed_M_STAR_MF}
		\vspace{-2mm}
	\end{figure}
	
	Fig. \ref{M_STAR_MF} illustrates the sum-rate versus the number of elements.
	The active RIS outperforms the STAR-RIS as $M$ increases, owing to its ability to directly amplify the signal. 
	Since a larger $M$ means that there are more elements operating in S mode, the gap between the MF-RIS and the STAR-RIS becomes negligible. 
	In contrast, maintaining self-sustainability for the self-sustainable RIS is costly.
   This further confirms that the proposed MF-RIS can effectively compensate for the performance loss caused by self-sustainability through full-space coverage and signal power enhancement.
	
	In Fig. \ref{Fixed_M_STAR_MF}, we plot the sum-rate versus the number of elements operating in S mode to exhibit the relationship between sum-rate maximization and energy harvesting maximization. 
	Here, we define $M_{\rm H}$ and $M_{\rm S}$ as the numbers of elements in H and S modes, respectively, satisfying $M_{\rm H}\!=\!\!M\!-\!\sum_{m\in\mathcal{M}}{\alpha_{m}}$ and $M_{\rm S}\!=\!\!\sum_{m\in\mathcal{M}}{\alpha_{m}}$.
    The sum-rate of the MF-RIS first increases and then decreases as $M_{\rm S}$ increases, which deviates from the common sense for passive RISs that more signal relay elements always benefit.
	This is because the trade-off between $M_{\rm S}$ and $M_{\rm H}$ at a fixed $M$ brings a trade-off between sum-rate and energy harvesting.
	Specifically, when $M_{\rm S}$ is small, the increase in $M_{\rm S}$ leads to a decrease in $M_{\rm H}$ and degrades the energy harvesting performance, but the relatively large $M_{\rm H}$ can harvest enough energy for the elements in S mode.
    Therefore, these signal relay elements can take advantage of reshaping the full-space wireless channels and mitigating their double attenuation to enhance signal reception.
     	\begin{figure}[t]
    	\centering
    	\includegraphics[width=3.05in]{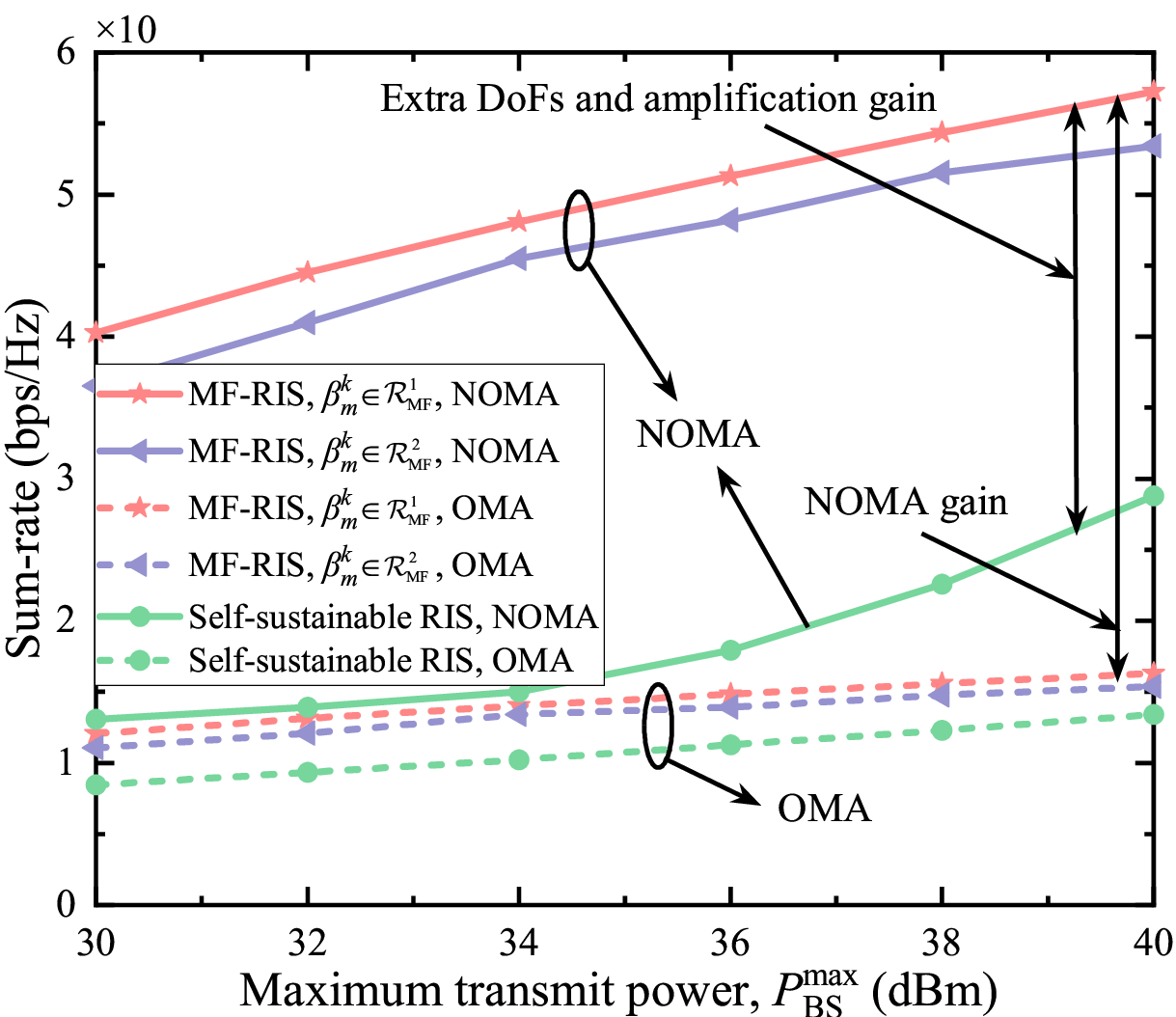}
    	\vspace{-2mm}
    	\caption{Sum-rate versus $P_{\rm BS}^{\max}$ under different schemes.}
    	\label{Pmax_OMA_NOMA}
    	\vspace{-2mm}
    \end{figure}
    
	Fig. \ref{Fixed_M_STAR_MF} shows that the sum-rate decreases as $M_{\rm S}$ increases after reaching the optimal value. 
	This is because that the decrease in $M_{\rm H}$ substantially restricts the energy that can be harvested, whereas the increase in $M_{\rm S}$ leads to higher circuit power consumption. 
	Consequently, the available amplification power at the MF-RIS is significantly reduced, making the MF-RIS suffer more from the increased $M_{\rm S}$.
	When $M_{\rm S}$ exceeds the maximum supportable value, the limited harvested energy may not even maintain self-sustainability, resulting in the failure of self-sustainable RISs.
	Since a larger $\beta_{\max}$ generates a greater output power, the optimal $M_{\rm S}$ for the scheme of $\beta_{\max}\!=\!16$ dB is larger than that of $\beta_{\max}\!=\!20$ dB.
	These results indicate that a flexible element allocation strategy is crucial for self-sustainable RIS schemes to balance the trade-off between sum-rate and energy harvesting. 
	A considerable performance gain is observed from the proposed MF-RIS and the self-sustainable RIS, verifying that the MF-RIS can better utilize the limited harvested power to enhance the sum-rate.

	In Fig. \ref{Pmax_OMA_NOMA}, we compare the achievable sum-rate of the considered NOMA and the conventional orthogonal multiple access (OMA) schemes.
	The MF-RIS and the self-sustainable RIS under NOMA yield a larger sum-rate value than their corresponding OMA schemes.
	Particularly, when $P_{\rm BS}^{\max}=40$ dBm, the NOMA systems assisted by the MF-RIS and the self-sustainable RIS attain $251$\% and $114.7$\% higher sum-rate gains than their OMA counterparts, respectively.
	The reason behind this is twofold:
	1) by serving all users within the same resource block, NOMA facilitates more flexible resource allocation to improve spectral efficiency;
     2) the location and coefficient design of RIS enable a smart NOMA operation by intelligently tuning the direction of users' channel vectors.
	
	\begin{figure}[t]
		\centering
		\includegraphics[width=3.05in]{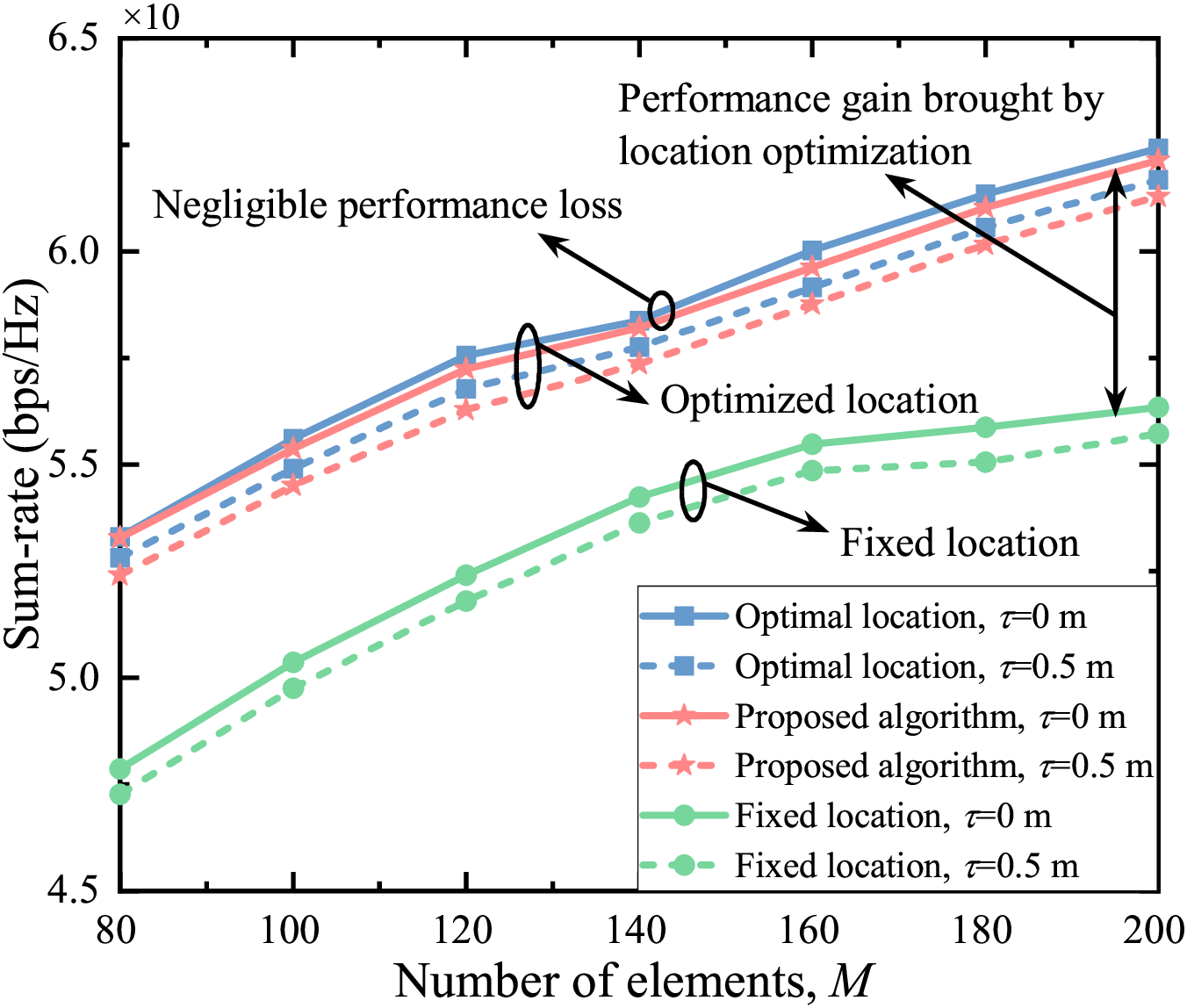}
		\vspace{-2mm}
		\caption{Sum-rate versus $M$ under different deployment strategies and different user location information.}
		\label{M_deployment}
		\vspace{-2mm}
	\end{figure}
	
	\begin{figure}[t]
		\centering
		\includegraphics[width=3.05in]{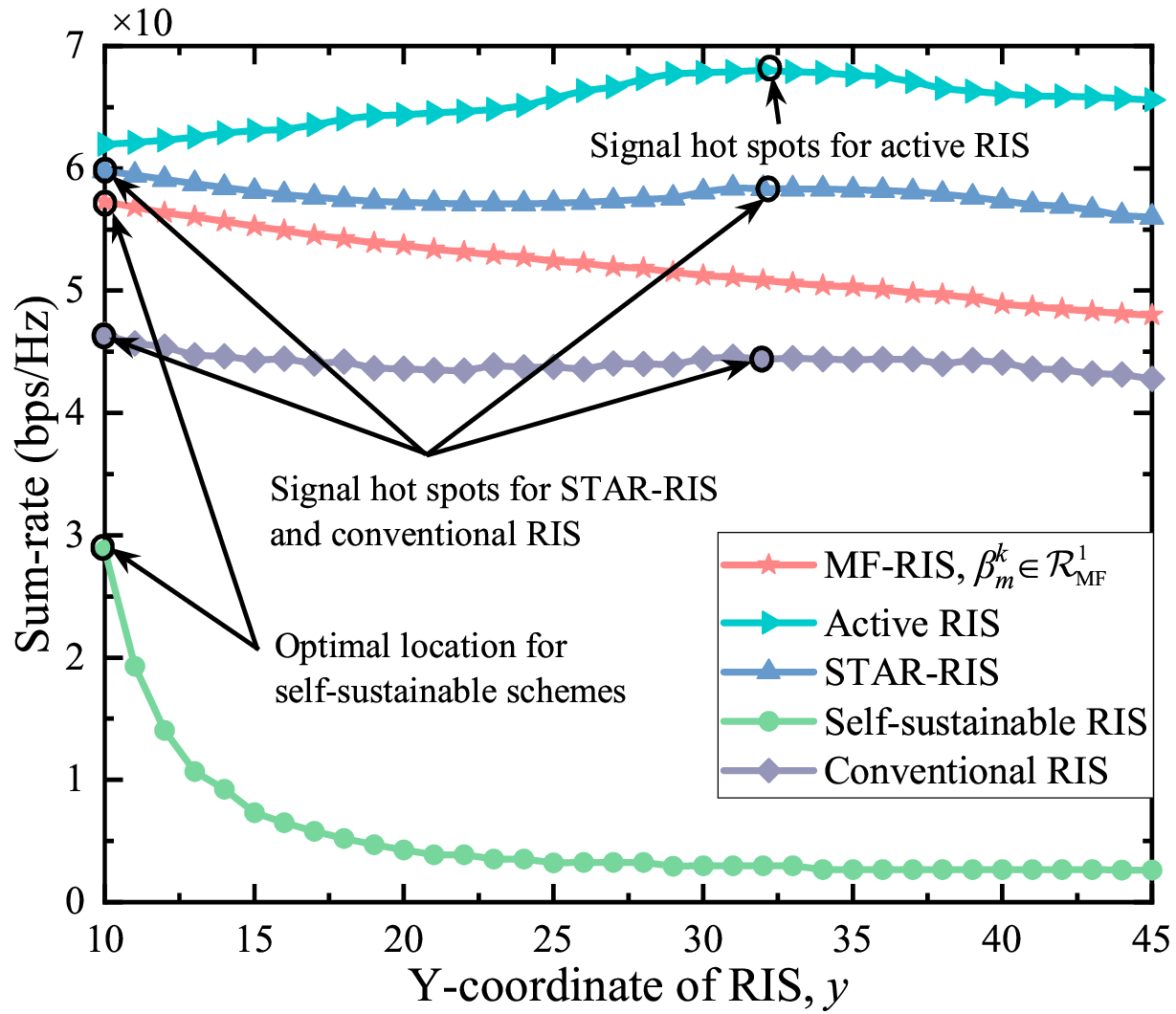}
		\vspace{-2mm}
		\caption{Sum-rate versus $y$ under different schemes.}
		\label{Y_location}
		\vspace{-2mm}
	\end{figure}
	
	Fig. \ref{M_deployment} depicts the achievable sum-rate versus the deployment strategy of RIS under different user location accuracy.
	We define $\tau$  as the location error of the user's actual location from the estimated location. 
	Here, ``Optimal location'' invokes exhaustive search to obtain the optimal RIS location, while ``Fixed location" fixes the RIS location at $\mathbf{w}=[5,20,10]^{\rm T}$ m.
	The ``Proposed algorithm'' significantly outperforms the ``Fixed location" approach, almost reaching the optimal performance.
	This is because optimizing the RIS location unleashes the full potential of RIS and NOMA, by providing a new DoF for their interplay.
	Another observation is that though the sum-rate achieved by the proposed algorithm decreases with increasing user location errors, the gaps between our algorithm and the benchmark schemes remain stable.
	This shows that the proposed algorithm maintains stable performance even at a low level of the location accuracy.
	
	In Fig. \ref{Y_location}, we study the impact of RIS location on sum-rate by varying the $Y$-coordinate of RIS. 
	For the STAR-RIS scheme, the sum-rate first decreases and then increases as $y$ increases, but decreases again after about $y\!=\!30$ m.
	This trend can be explained as follows.
	Since the channel gain is a decreasing function of the link distance, the STAR-RIS deployed in the vicinity of the BS or users creates signal hot spots.
	However, when the STAR-RIS is located at the middle between the BS and users, both the BS-RIS and RIS-user links experience severe signal attenuation.
	In contrast, the active RIS is less affected by double attenuation as it provides more amplification gain to compensate for path loss when moving away from the BS.
    The inflection point occurs when deploying the STAR-RIS and active RIS near the user closer to the BS, which allows exploiting the channel gain differences among multiple NOMA users.
	As for self-sustainable RIS schemes, the sum-rate decreases as the RIS moves away from the BS. The optimal deployment location is closest to the BS within the allowed range. 
	To maintain the balance between the energy supply and demand, both the MF-RIS and the self-sustainable RIS need to allocate more elements for energy harvesting when they are far away from the BS, resulting in fewer elements to relay signals.
	The above observations show that conventional RISs obtain good performance when they are deployed near the transmitter or receiver, but for self-sustainable RISs, it would be better to deploy closer to the transmitter.
	
	\section{Conclusion}\label{Conclusion}
	In this paper, we proposed an MF-RIS architecture enabling simultaneous signal reflection, refraction, amplification, and energy harvesting.
	The differences between the proposed MF-RIS and existing SF- and DF-RIS were first discussed from the perspective of the signal model.
	Next, we investigated the sum-rate maximization problem in an MF-RIS-aided NOMA network and the resulting MINLP problem was efficiently solved by an iterative algorithm.
	Numerical results provided useful insights for practical system design, which, in particular, are
	1) compared to the conventional passive RIS and self-sustainable passive RIS, the proposed MF-RIS attain $23.4$\% and $98.8$\% performance gains, respectively, by integrating multiple functions on one surface;
	and 2) deploying MF-RIS closer to the transmitter side facilitates energy harvesting and therefore brings a higher sum-rate gain.	
	Several interesting directions to pursue include:
	\begin{itemize}
		\item {\emph {Practical implementation:}}
		Compared to the prototype design of existing SF- and DF-RIS, the proposed MF-RIS faces new implementation challenges. For example, how to embed circuits that perform signal reflection, refraction, amplification, and energy harvesting functions into a limited substrate simultaneously, and how to balance the efficiency of these circuits.
		
		\item {\emph {High-accuracy channel estimation:}} The proposed MF-RIS requires more pilot overhead than existing SF-RIS to estimate the reflection and refraction channels. 
		Although simultaneously estimating all channels reduces the overhead, how to achieve fast and high-accuracy channel estimation requires further research.	
			
		\item {\emph {Low-complexity deployment:}} 
		To make the proposed MF-RIS easy to deploy in practical systems, we can group the elements and set the same reflective/refractive amplitude for each group. Nevertheless, it remains open how to group the elements during practical deployment to attain desired performance with low complexity.
	\end{itemize}

	\appendices
	
	\section{Proof of Lemma \ref{lemma-2}} \label{proof_of_lemma_2}
	 Based on the rate expression in (\ref{R_kj}) and the equivalently combined channel gain $\gamma_{kj}$, we obtain
	 \begin{align}
	 	\bar{p}_{kj}=r_{kj}^{\min}\left(\sum\nolimits_{i=j+1}^{J_k}\bar{p}_{ki}+\frac{1}{\gamma_{kj}}\right).
	 \end{align}
	According to \cite[Proposition 1]{Power-SPL}, the minimum transmit power is derived as
	\begin{align}
		\nonumber
		\sum\nolimits_{j\in\mathcal{J}_k}\bar{p}_{kj}&=\sum\nolimits_{j\in\mathcal{J}_k}r_{kj}^{\min}\left(\sum\nolimits_{i=j+1}^{J_k}\bar{p}_{ki}+\frac{1}{\gamma_{kj}}\right)\\
		&=\sum\nolimits_{j\in\mathcal{J}_k} \left(\prod\nolimits_{i=1}^{j-1}\left(r_{ki}^{\min}+1\right)\right)\frac{r_{kj}^{\min}}{\gamma_{kj}}.
	\end{align}
	Thus, in order to make problem (\ref{P_p-1}) feasible, power allocation coefficients should satisfy (\ref{p_feasible}). 
	
	\section{Proof of Proposition \ref{proposition-1}} \label{proof_of_proposition_1}	
	Following the rate-splitting principle and defining the function $\Psi(x)=\log_2(1+x)$, we rewrite the achievable rate expression $R_{j\to j}^k$ as follows\cite{Power-TIT}:
	\begin{align}
\nonumber
		 R_{j\to j}^k\!&\!\overset{(a)}{=}\!\Psi\Big[\frac{\gamma_{kj}p_{kj}}{\gamma_{kj}\sum\nolimits_{i=j+1}^{J_k}{p}_{ki}+1}\Big] 
		\!\overset{(b)}{=}\!\Psi\Big[\frac{\gamma_{kj}(\bar{p}_{kj}+\triangle p_{kj})}{\gamma_{kj}\sum\nolimits_{i=j+1}^{J_k}{p}_{ki}+1}\Big]\\
		\nonumber
		&\!\overset{(c)}{=}\!\Psi\Big[\frac{ \bar{p}_{kj}+r_{kj}^{\min}\sum\nolimits_{i=j+1}^{J_k}\triangle p_{ki}}{\sum\nolimits_{i=j+1}^{J_k}p_{ki}+\frac{1}{\gamma_{kj}}}\Big] \\
        \nonumber
		& ~~~\!+\!\Psi\Big[\frac{ \triangle {p}_{kj}-r_{kj}^{\min} \sum\nolimits_{i=j+1}^{J_k}\triangle p_{ki} }{ 
			\bar{p}_{kj}+r_{kj}^{\min}\sum\nolimits_{i=j+1}^{J_k}\triangle p_{ki} + \sum\nolimits_{i=j+1}^{J_k} p_{ki} +\frac{1}{ \gamma_{kj}} }\Big]
		\\
		\nonumber
		&\!\overset{(d)}{=}\!\Psi\Big[\underbrace{\frac{ \bar{p}_{kj}+r_{kj}^{\min} \sum\nolimits_{i=j+1}^{J_k}\triangle p_{ki} }{\sum\nolimits_{i=j+1}^{J_k}\bar{p}_{ki}+\sum\nolimits_{i=j+1}^{J_k}\triangle p_{ki} + \frac{1}{\gamma_{kj}}}}_{r_{kj}^{\min}}\Big] \\
		\label{Appendix-R-C}
		&~~ +\Psi\Big[\underbrace{\frac{ \triangle {p}_{kj} -  r_{kj}^{\min}\sum\nolimits_{i=j+1}^{J_k}\triangle p_{ki} }{\sum\nolimits_{i=j}^{J_k}\bar{p}_{ki}  +  (1  +  r_{kj}^{\min})\sum\nolimits_{i=j+1}^{J_k}\triangle p_{ki}   + \frac{1}{\gamma_{kj}} }}_{\triangle r_{kj}^{\min}}\Big],  
	\end{align}
    where $(b)$ follows from $p_{kj}=\bar{p}_{kj}+\triangle p_{kj}$, $(c)$ holds due to the rate-splitting property\cite{Power-TIT}, i.e., $\Psi(\frac{x+y}{z})=\Psi(\frac{x}{z})+\Psi(\frac{y}{x+z})$, and $(d)$ is due to the fact that the following equation holds:
    \begin{align}
    	\nonumber
    	&\bar{p}_{kj}+r_{kj}^{\min}\sum\nolimits_{i=j+1}^{J_k}\triangle  p_{ki} \\
    	&=r_{kj}^{\min}\left(\sum\nolimits_{i=j+1}^{J_k}\bar{p}_{ki}+\frac{1}{\gamma_{kj}}\right)+r_{kj}^{\min}\sum\nolimits_{i=j+1}^{J_k}\triangle p_{ki}.
    \end{align}
    
    To simplify the expression of $\triangle r_{kj}^{\min}$, we define
    \begin{subequations}
    	\label{p_deltap}
    	  \begin{align}
    		\label{p_deltap-1}
    		&\!\!\!\triangle \hat{p}_{kj} \! =\! \big(\triangle p_{kj} \! - \! r_{kj}^{\min} \sum\nolimits_{i=j+1}^{J_k} \!\!\triangle p_{ki} \big)\prod\nolimits_{i=1}^{j-1}(1\! + \! r_{ki}^{\min}), \!\!\!\!\!\! \\
    		\label{p_deltap-2}
    		&\!\!\!\hat{p}_{kj}\!=\!\big(\!\sum\nolimits_{i=j}^{J_k}\bar{p}_{ki} \! + \! \frac{1}{\gamma_{kj}}\big)\prod\nolimits_{i=1}^{j-1}(1 \! + \! r_{ki}^{\min}).\!\!\!\!\!\!
    	\end{align}
    \end{subequations}
  Based on  (\ref{p_deltap}), we then obtain the following equations:
   \begin{subequations}
    \begin{align}
    	&\sum\nolimits_{i=j+1}^{J_k}\triangle \hat{p}_{ki}=\prod\nolimits_{i=1}^{j}(1+r_{ki}^{\min}) \sum\nolimits_{i=j+1}^{J_k}\triangle p_{ki}, \\
    	&\triangle r_{kj}^{\min}= \frac{ \triangle \hat{p}_{kj} }{\hat{p}_{kj} + \sum_{i=j+1}^{J_k} \triangle \hat{p}_{ki}}.
    \end{align}
 \end{subequations}
 As a result, problem (\ref{P_p-1}) is reformulated as
	\begin{subequations}
		\label{P_p-appendix}
		\begin{eqnarray}
			\label{P_p-appendix-function}
			& \underset{\triangle \hat{p}_{kj} }{\max}  & \sum\nolimits_{j\in\mathcal{J}_k}\Psi(r_{kj}^{\min})+\sum\nolimits_{j\in\mathcal{J}_k}\Psi (\triangle r_{kj}^{\min}) \\
			& \operatorname{s.t.} & \sum\nolimits_{j\in\mathcal{J}_k} \triangle \hat{p}_{kj} =1-\sum\nolimits_{j\in\mathcal{J}_k}\bar{p}_{kj},
		\end{eqnarray}
	\end{subequations}
	where $\sum\nolimits_{j\in\mathcal{J}_k}\Psi(r_{kj}^{\min})$ is a constant that does not affect the optimality of (\ref{P_p-appendix}).
	Since the inequality $\hat{p}_{kj}\geq \hat{p}_{k(j+1)}$ holds and $\Psi(\triangle r_{kj}^{\min})$ increases with $\triangle {r}_{kj}^{\min}$, the optimal solution of (\ref{P_p-appendix}) is to allocate all the excess power to user $U_{kJ_k}$ with the best equivalent channel gain, i.e., 
	\begin{align}
		\!\!\!\triangle \hat{p}_{kJ_k}=1-\sum\nolimits_{j\in\mathcal{J}_k}\bar{p}_{kj}, ~~\triangle \hat{p}_{kj}=0, ~ \forall j\in\{\mathcal{J}_k/J_k\}.\!\!\!
	\end{align}
	Finally, the optimal power allocation coefficients and the corresponding
	objective value of problem (\ref{P_p-1}) are obtained as (\ref{optimal_p}) and (\ref{optimal_R_j}), respectively.

	\section{Proof of Constraint (\ref{C_passive_rank_2-relax})}	 \label{proof_of_MF-RIS_coefficient}
	We apply the penalty-based method to handle constraint $\eta_m^k=\alpha_{m}^2\beta_{m}^k$.
	Note that if we directly add it as a penalty term into the objective function (\ref{P_passive-1-objective}), (\ref{P_passive-1-objective}) will become $\sum\nolimits_{k\in\mathcal{K}}\sum\nolimits_{j\in\mathcal{J}_k}Q_{kj}$ $-\rho \sum\nolimits_{k\in\mathcal{K}}\sum\nolimits_{m\in\mathcal{M}} (\alpha_{m}^2\beta_{m}^k-\eta_m^k)$, where $\rho>0$ denotes the penalty factor.
	The resultant objective function is non-concave due to the term $\alpha_{m}^2\beta_{m}^k$.
	To this end, we replace it with its CUB \cite{Wen-DF-RIS}.
	Define the functions $g_1(\alpha_{m},\beta_{m}^k)=\alpha_{m}^2\beta_{m}^k$ and $g_2(\alpha_{m},\beta_{m}^k)=\frac{c_m^k}{2}\alpha_{m}^4+\frac{(\beta_{m}^k)^2}{2c_m^k}$.
	Then it is easy to check that $g_2(\alpha_{m},\beta_{m}^k)$ is a convex overestimate of $g_1(\alpha_{m},\beta_{m}^k)$ for $c_m^k>0$.  
	Moreover, when $c_m^k=\frac{\beta_{m}^k}{\alpha_{m}^2}$, the equations $g_1(\alpha_{m},\beta_{m}^k)=g_2(\alpha_{m},\beta_{m}^k)$ and $\nabla g_1(\alpha_{m},\beta_{m}^k)=\nabla g_2(\alpha_{m},\beta_{m}^k)$ hold, where $\nabla g_1(\alpha_{m},\beta_{m}^k)$ and $\nabla g_2(\alpha_{m},\beta_{m}^k)$ are the gradients of $g_1(\alpha_{m},\beta_{m}^k)$ and $g_2(\alpha_{m},\beta_{m}^k)$, respectively.
	
	\section{Terms Introduced in Constraint (\ref{C_appendix-D-ABC})} \label{Parameters introduced in constraint}
	The terms introduced in constraint (\ref{C_appendix-D-ABC}) that are unrelated to the optimization variable in the $i$-th iteration are given by
	\begin{subequations}
		\label{W_independent}
		\begin{align}
			&\mathbf{D}_{kj}=\mathbf{E}_{kj}\mathbf{F}_k\mathbf{E}_{kj}^{\rm H},
			~\bar{\mathbf{D}}_{kj}=\mathbf{E}_{kj}\mathbf{F}_{\bar{k}}\mathbf{E}_{kj}^{\rm H}, \\
			&\mathbf{E}_{kj}=\big[\mathbf{h}_{kj}, h_0\mathbf{P}_k^{\rm H}\hat{\mathbf{g}}_{kj}^{(i-1)}\big]^{\rm H}, 
			~\mathbf{P}_{k}=\boldsymbol{\Theta}_k\hat{\mathbf{H}}^{(i-1)},\\
			&\mathcal{W}_1=h_0\sigma_s^2\lVert{(\hat{\mathbf{g}}_{kj}^{(i-1)})}^{\mathrm H}\boldsymbol{\Theta}_{k}\lVert^2, \\
			&\mathcal{W}_2=\frac{\bar{\mathcal{W}}-\sigma_s^2\sum_{k\in\mathcal{K}}\lVert \boldsymbol{\Theta}_k\lVert_F^2}{h_0\sum\nolimits_{k\in\mathcal{K}} \big( \sum_{k'\in\mathcal{K}}{\rm  Tr} (\mathbf{P}_k\mathbf{F}_{k'}\mathbf{P}_k^{\rm H}) \big)}, \\
			&\mathcal{W}_3=\frac{\zeta_m-\sigma_s^2(1-\alpha_{m})}{h_0(1-\alpha_{m}) \sum\nolimits_{k\in\mathcal{K}} \operatorname{Tr}\big( \bar{\mathbf{T}}_m\hat{\mathbf{H}}^{(i-1)}\mathbf{F}_k{(\hat{\mathbf{H}}^{(i-1)})}^{\rm H}\bar{\mathbf{T}}_m^{\rm H}\big)}.
		\end{align}
	\end{subequations}
	
	\section{Proof of Constraint (\ref{C_appendix-D-ABC})} \label{proof_of_lemma_3}
	We define the slack variable set $\Delta_2=\{t, t_{kj}, \bar{t}_{kj}, e_{kj},v,$ $\bar{v}, r_{kj},\bar{r}_{kj}, s_{kj}\}$ as
	\begin{subequations}
		\begin{align}
			&\!\!\!\!\!\! \!\!\! t\!=\!d_{bs}^{-\frac{\kappa_{0}}{2}},
			{t}_{kj}\!=\!d_{skj}^{-\frac{\kappa_{2}}{2}},
			\bar{t}_{kj}\!=\! t{t}_{kj}, 
			 e_{kj}\!=\!d_{skj}^{-\kappa_{2}}, \\
			&\!\!\!\!\!\! \!\!\! v\!=\!\bar{v}=d_{bs}^{-\kappa_{0}},
		    r_{kj}\!=\!\bar{r}_{kj}\!=\!\bar{\mathbf{d}}_{kj}^{\rm T}\mathbf{D}_{kj}\bar{\mathbf{d}}_{kj},
			s_{kj}\!=\!\bar{\mathbf{d}}_{kj}^{\rm T}\bar{\mathbf{D}}_{kj}\bar{\mathbf{d}}_{kj}.\!\!\!\!\!\!\!\!\!
		\end{align}
	\end{subequations}
	Constraints (\ref{C_AB_w}) and (\ref{C_C_w}) are then rewritten as
	\begin{subequations}
	\begin{align}  
		\label{C_slack variable-1}
		&t\leq  d_{bs}^{-\frac{\kappa_{0}}{2}},
		~~t_{kj}\leq d_{skj}^{-\frac{\kappa_{2}}{2}}, 
		~~e_{kj}\geq d_{skj}^{-\kappa_{2}}, \\
		\label{C_slack variable-2}
		&v\geq d_{bs}^{-\kappa_{0}}, 
		~~\bar{v}\leq d_{bs}^{-\kappa_{0}},
		~~r_{kj}\leq \bar{\mathbf{d}}_{kj}^{\rm T}\mathbf{D}_{kj}\bar{\mathbf{d}}_{kj}, \\
		\label{Appendix-D-1}
		&\bar{t}_{kj}\leq tt_{kj}, 
		~~\bar{r}_{kj}\geq \bar{\mathbf{d}}_{kj}^{\rm T}{\mathbf{D}}_{kj}\bar{\mathbf{d}}_{kj},
		~~s_{kj}\geq \bar{\mathbf{d}}_{kj}^{\rm T}\bar{\mathbf{D}}_{kj}\bar{\mathbf{d}}_{kj},\\
		&A_{kj}^{-1}\leq r_{kj},~B_{kj}\geq s_{kj}+e_{kj}\mathcal{W}_{1}+\sigma_u^2, \\
		\label{Appendix-D-2}
		& C_{kj}\geq \bar{r}_{kj} P_{kj}+B_{kj}, 
		~~v \leq \mathcal{W}_2, 
		~~\bar{v}\geq \mathcal{W}_3.
	\end{align}
   \end{subequations}
     
    Since the constraints in (\ref{C_slack variable-1}) and (\ref{C_slack variable-2}) are still non-convex, we apply the SCA method to deal with them.
	Specifically, by exploiting the first-order Taylor expansion of $\bar{\mathbf{d}}_{kj}^{\rm T}\mathbf{D}_{kj}\bar{\mathbf{d}}_{kj}$ at the given point $\{\bar{\mathbf{d}}_{kj}^{(\ell)}\}$, constraint $r_{kj}\leq \bar{\mathbf{d}}_{kj}^{\rm T}\mathbf{D}_{kj}\bar{\mathbf{d}}_{kj}$ is recast as the following convex one:
	\begin{align}
	\label{Appendix-D-3}
	r_{kj}\leq  -(\bar{\mathbf{d}}_{kj}^{(\ell)})^{\rm T}\mathbf{D}_{kj}\bar{\mathbf{d}}_{kj}^{(\ell)}
	+2 \Re\big((\bar{\mathbf{d}}_{kj}^{(\ell)})^{\rm T}\mathbf{D}_{kj}\bar{\mathbf{d}}_{kj}\big).
	\end{align}
    To facilitate the derivation of the other constraints in (\ref{C_slack variable-1}) and (\ref{C_slack variable-2}), we rewrite them as follows:
	\begin{subequations}
		\label{Appendix-FTS}
		\begin{eqnarray}
			&\!\!\!\!\!\!\!\!\!\!\!\!g(x,y,z)\!+\!g(x_b,y_b,z_b)\! -\!2x_bx\! - \! 2y_by
			\! - \! 2z_bz\! -\! t^{-\frac{4}{\kappa_{0}}} \! \leq \! 0,\\
			&\!\!\!\!\!\!\!\!\!\!\!\! \!\!\! \!\!\! g(x,y,z)\! + \! g( x_{kj}, y_{kj}, 0) \!-\! 2x_{kj}x\! -\! 2y_{kj}y
			\! - \! t_{kj}^{-\frac{4}{\kappa_{2}}}\! \leq \! 0, \\
			& \!\!\!\!\!\!\!\!\!\!\!\!\!\! -g(x,y,z)\! - \! g( x_{kj}, y_{kj}, 0) \!+\! 2x_{kj}x\! +\! 2y_{kj}y
			\! +\! e_{kj}^{-\frac{2}{\kappa_{2}}}\! \leq \! 0,\\
			&\!\!\!\!\!\!\!\!\!\!\!\! - \! g(x,y,z) \! \! - \!\! g(x_b,y_b,z_b)\! + \! 2x_{b}x\! \! + \! \! 2y_{b}y
			\! + \! 2z_bz\! + \! v^{-\frac{2}{\kappa_{0}}}\! \leq\!  0, \\
			&\!\!\!\!\!\!\!\!\!\!\!\!\! g(x,y,z)\! + \! g(x_b,y_b,z_b) \! -\! 2x_{b}x\! - \! 2y_{b}y
			\! - \! 2z_bz\! -\! \bar{v}^{-\frac{2}{\kappa_{0}}}\! \leq \!0,
		\end{eqnarray}
	\end{subequations}
    where the function $g(a,b,c)$ is defined as $g(a,b,c)= a^2+b^2+c^2$.
	The existence of non-convex terms $-t^{-\frac{4}{\kappa_{0}}}$, $-t_{kj}^{-\frac{4}{\kappa_{2}}}$, $-{\bar{v}}^{-\frac{2}{\kappa_{0}}}$, $-x^2$, $-y^2$, and $-z^2$ makes (\ref{Appendix-FTS}) non-convex.
	By replacing these terms with their respective convex first-order Taylor expansions, we obtain the following convex ones:
   \begin{subequations} 
   	\label{P_w-C-SCA-appendix}
	\begin{eqnarray}
	\nonumber
	& g(x,y,z)  +  g(x_b,y_b,z_b)  - 2x_bx -  2y_by - 2z_bz  \\
   &\!\!\!\!\!\!\!\!\!\!\! \!\!\!\!\!\!\!\!\!\!\!\!\!\!\!\!\!\!\!\!\!\!\!\!\!\!\!\!\!\!\!\!\!\!\!\!\!\!\!\!\!\!\!\!\!\!\!\!\!\!\!\!\!\!\!\!\!\!\!\!\!\!\! + f(t, -\frac{4}{\kappa_{0}}) \leq  0,\\
    \nonumber
	&\!\!\! \!\!\!\!\!\! \!\!\! g(x,y,z)  + g( x_{kj}, y_{kj}, 0)  - 2x_{kj}x  - 2y_{kj}y  \\
	&\!\!\!\!\!\!\!\!\!\!\!\!\!\!\!\!\!\!\!\!\!\!\!\!\!\!\!\!\!\!\!\!\!\!\!\!\!\!\!\!\!\!\!\!\!\!\!\!\!\!\!\!\!\!\!\!\!\!\!\!\!\!\!\!\!\!\!\!\!\!\! + f(t_{kj},  -\frac{4}{\kappa_{2}})  \leq  0, \\
	\nonumber
	&\!\!\!\!\!\!\!\!\!\!\!\!  f(x, 2)  +  f(y, 2)  + f(z,2) - g( x_{kj}, y_{kj}, 0)  \\
	&\!\!\!\!\!\!\!\!\!\!\!\!\!\!\!\!\!\!\!\!\!\!\!\!\!\!\!\!\!\!\!\!\!\!\!\!\!\!\!\!\!\! +  2x_{kj}x  +   2y_{kj}y   + e_{kj}^{-\frac{2}{\kappa_{2}}}  \leq  0,\\
	\nonumber
	&\!\!\!\!\!\!\!\!\!\!\!\!\!\!\! f(x, 2) +  f(y,2)  +  f(z, 2) - g( x_b, y_{b}, z_b)   \\
	& \!\!\!\!\!\!\!\!\!\!\!\!\!\!\!\!\!\! \!\!\!\!\!\! \!\!\!\! +  2x_{b}x + 2y_{b}y+2z_bz+ v^{-\frac{2}{\kappa_{0}}}\leq 0,\\
	\nonumber
	& \!\!\!\!\!\!\!\!\!\!\!\!\!\!\!\! g(x,y,z)  + g(x_b,y_b,z_b) - 2x_bx  - 2y_by\\
	& \!\!\!\!\!\!\!\!\!\!\!\!\!\!\!\!\!\!\!\!\!\!\!\!\!\!\!\!\!\!\!\!\!\!\!\!\!\!\!\!\!\!\!\!\!\!\!\!\!\!\!\!\!\!\!\! - 2z_bz + f(\bar{v}, -\frac{2}{\kappa_{0}})  \leq 0,
  \end{eqnarray}
  \end{subequations}
 where $f(p,q)=-{(p^{(\ell)})}^{q}-q{(p^{(\ell)})}^{q-1}(p-p^{(\ell)})$ is the first-order Taylor expansion of $-p^q$ at the given point $\{p^{(\ell)}\}$.

\end{document}